\documentclass[twocolumn,showpacs,aps,prc,superscriptaddress]{revtex4}

\usepackage[dvips]{graphicx}

\begin{document}

\title{Backward electroproduction of $\pi^0$ mesons on protons in the region
of nucleon resonances at four momentum transfer squared $Q^2 = 1.0$~GeV$^2$}

\author{G.~Laveissi\`{e}re}
\affiliation{Universit\'{e} Blaise Pascal/IN2P3, F-63177 Aubi\`{e}re, France}
\author{N.~Degrande}
\affiliation{University of Gent, B-9000 Gent, Belgium}
\author{S.~Jaminion}
\affiliation{Universit\'{e} Blaise Pascal/IN2P3, F-63177 Aubi\`{e}re, France}
\author{C.~Jutier}
\affiliation{Universit\'{e} Blaise Pascal/IN2P3, F-63177 Aubi\`{e}re, France}
\affiliation{Old Dominion University, Norfolk, VA 23529}
\author{L.~Todor}
\affiliation{Old Dominion University, Norfolk, VA 23529}
\author{R.~Di Salvo}
\affiliation{Universit\'{e} Blaise Pascal/IN2P3, F-63177 Aubi\`{e}re, France}
\author{L.~Van Hoorebeke}
\affiliation{University of Gent, B-9000 Gent, Belgium}
\author{L.C.~Alexa}
\affiliation{University of Regina, Regina, SK S4S OA2, Canada}
\author{B.D.~Anderson}
\affiliation{Kent State University, Kent OH 44242}
\author{K.A.~Aniol}
\affiliation{California State University, Los Angeles, CA 90032}
% ...
\collaboration{The Jefferson Lab Hall A Collaboration}
\noaffiliation

\makeatletter
\global\@specialpagefalse
\def\@oddhead{\hfill {G. Laveissiere {\it et al.}} {$\pi^0$ electro production}}
\let\@evenhead\@oddhead
\def\@oddfoot{\reset@font\rm\hfill \thepage\hfill
%\ifnum\c@page=1
%\llap{\protect\copyright{} }
%\fi
} \let\@evenfoot\@oddfoot
\makeatother

\begin{abstract}
Exclusive electroproduction of $\pi^0$ mesons on protons in the backward hemisphere
has been studied at $Q^2$ = 1.0 GeV$^2$ by detecting protons in the forward direction
in coincidence with scattered electrons from the 4~GeV electron beam in Jefferson
Lab's Hall A. The data span the range of the total ($\gamma^*p$) center-of-mass
energy $W$ from the pion production threshold to $W = 2.0$~GeV. The differential
cross sections $\sigma_{\mbox{\tiny T}}+\epsilon\ \sigma_{\mbox{\tiny L}}$,
$\sigma_{\mbox{\tiny TL}}$, and $\sigma_{\mbox{\tiny TT}}$ were separated from the
azimuthal distribution and are presented together with the
MAID~\cite{Drechsel:1998hk,Kamalov:2001yi,Tiator:2003uu} and
SAID~\cite{Arndt:1996ak,Arndt:2002xv,Arndt:2003zk} parametrizations.
\end{abstract}

\pacs{14.20.Dh,14.20.Gk}

\maketitle

%%%%% SECTION I %%%%%%%%%%%%%%%%%%%%%%%%%%%%%%%%%%%%%%%%%%%%%%%%%%%%%%%%%%%%%%%%%
\section{Introduction}
\label{sec:introduction}
%%%%%%%%%%%%%%%%%%%%%%%%%%%%%%%%%%%%%%%%%%%%%%%%%%%%%%%%%%%%%%%%%%%%%%%%%%%%%%%%%
The present experiment~\cite{Bertin:1993} exploits the attractive opportunity
to investigate a number of resonance states by detecting their decay
into two channels of very similar kinematics, but remarkably different final-state
interaction (FSI) couplings. They are:
\begin{equation}
e^- + p \to e^- + p + \pi^0
\label{react1}
\end{equation}
and
\begin{equation}
e^- + p \to e^- + p + \gamma.
\label{react2}
\end{equation}

The intermediate resonant state decays via the strong interaction in
reaction~(\ref{react1}), and via the electromagnetic interaction in
reaction~(\ref{react2}). However, one can employ an identical technique
for detecting two of the three outgoing particles for both reactions,
namely detection of the scattered electron and proton in coincidence.
This results in a greater precision for the relative cross sections
of the two reactions than for either cross section alone.

A comparison of reactions~(\ref{react1}) and~(\ref{react2})
may be beneficial in addressing the problem of the ``missing'' resonances.
The Constituent Quark Model (CQM)~\cite{Isgur:1979wd} predicts several
positive parity states at $W>1.6$ GeV that have not been observed
~\cite{Koniuk:1980vw,Koniuk:1980vy,Koniuk:1982ej,Capstick:1994kb,Bijker:1996ii}.
It is conjectured that these states couple relatively weakly to the $\pi N$
channel which has dominated (either in the initial or final state) most
of the experimental work to date.
The two reactions~(\ref{react1}) and~(\ref{react2}) provide therefore a
potentially very different sensitivity to the missing resonances.

Closely related to reactions~(\ref{react1}) and~(\ref{react2}) is the
process of deep inelastic electron scattering, which is generally
analysed in terms of parton rather than baryon resonance degrees of
freedom. However, the phenomenon of quark-hadron duality illustrates
the interplay of these two frameworks at modest $Q^2$ and
$W$~\cite{Niculescu:2000tk,Niculescu:2000tj}. Another motivation for the
present study is to explore the exclusive reactions~(\ref{react1})
and~(\ref{react2}) in the high energy limit, where current quark
degrees of freedom may play as important a role as resonance, or
constituent quark, degrees of freedom.

In the absence of a theoretical approach based on fundamental
principles, one has to rely on experimental input and use
phenomenological models. In the
region of the first resonance, $\Delta(1232)$, many models are
well-developed and are successful in describing the resonance
spectrum quantitatively: MAID~\cite{Drechsel:1998hk,Kamalov:2001yi,Tiator:2003uu},
SAID~\cite{Arndt:1996ak,Arndt:2002xv,Arndt:2003zk},
and others~\cite{Davidson:1991xz,Aznaurian:1998sb,Aznauryan:2002gd}.
One finds a substantial increase in  uncertainties for masses and
hadronic and electromagnetic couplings of higher resonances where
resonant and non-resonant channels compete.
An increase in the total center-of-mass (CM) energy $W$ is followed
by an increase in the number of coupled channels, which have to be
related via unitarity.
At this point, even the use of all available data on the resonance
production cannot resolve the difficulties of the model approaches
in particular for $W > 1.6$~GeV.
To constrain these hadronic and EM couplings, there is currently an
intensive world-wide effort to simultaneously study all decay
channels produced in photo- and electromagnetic excitation of the
nucleon (see for example~\cite{Merkel:2001qg,Ahrens:2003na} or~\cite{Burkert:2002nr}).

\begin{figure}
\includegraphics[width=6cm]{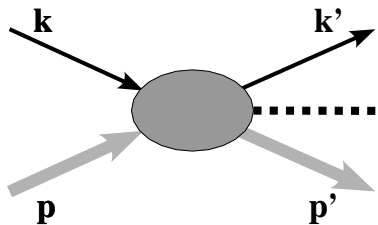}
\caption{\label{kinematic}Definition of kinematic variables for reaction~(\ref{react1}).
         Thin lines represent the incident and outgoing electrons, and thick
         lines correspond to the target and recoil protons. The dashed line
         stands for the produced neutral pion.}
\end{figure}

The results for reaction~(\ref{react2}) will be presented in another paper.
In this paper we present cross-section measurements of reaction~(\ref{react1})
made in Hall~A of the Thomas Jefferson National Accelerator Facility (JLab)
at an incident electron energy 4~GeV and fixed four-momentum transfer squared
$Q^2$ = 1.0~GeV$^2$.
The scattered electron and proton (momenta $k'$ and $p'$ respectively) are detected
at laboratory angles $\theta_e$ and $\theta_p$, and the neutral pion is
reconstructed using a missing-mass technique.
The missing mass squared is expressed as $M_X^2= (k+p-k'-p')^2$ where
$k$ and $p$ are the momentum of the initial electron and proton, respectively.
The relevant kinematical variables are shown in Fig.~\ref{kinematic}.

The kinematics were further restricted to forward detection (relative to the
virtual photon momentum vector) of the recoil proton (backward CM $\pi^0$ emission).
This reaction has been studied previously at the NINA electron
synchrotron at a beam energy of 4~GeV~\cite{Siddle:1971ug},
at DESY at 2.7 and 3.2~GeV~\cite{Galster:1972rh,Alder:1972di,Alder:1976xt,Albrecht:1970xb,Albrecht:1971rv}
and, recently, in Hall~C~\cite{Frolov:1998pw} and Hall~B\cite{Joo:2001tw,Joo:2003uc,Biselli:2003ym}
experiments at JLab.

Our results will be presented as the conventional center of mass photo-production
cross section, where the photon flux factor $\Gamma$ (Hand convention) is introduced
in the one-photon-exchange approximation:
\begin{equation}
\frac{d^5\sigma}{dk'd\Omega_ed\Omega_{\pi}^\ast}=
\Gamma\ \frac{d^2\sigma}{d\Omega_{\pi}^\ast},
\end{equation}
\begin{equation}
\Gamma=\frac{\alpha}{2\pi^2}\ \frac{k'}{k}\ \frac{W^2-{\rm M}_p^2}{2{\rm M}_p\cdot Q^2}
\ \frac{1}{1-\epsilon},
\end{equation}
where $\Omega_e$ is the differential solid angle for the scattered electron
in the LAB frame, $\Omega_{\pi}^\ast$ is the differential solid angle for the
proton in the final pion-proton CM frame, $M_p$ is the proton mass,
$\alpha$ is the fine-structure constant, and $\epsilon$ is the virtual
photon polarization:
\begin{equation}
\epsilon=\frac{1}{1+2\frac{(\vec{k}-\vec{k'})^2}{Q^2}tan^2(\frac{\theta_e}{2})}.
\end{equation}

In the following, $\theta^*$ is defined as the polar angle between
the virtual photon and the pion in the pion-proton
center of mass system.
$\phi$ is the azimuthal angle between the leptonic and the hadronic planes
($\phi$ is taken equal to 0 when the pion is emitted in the
half plane containing the outgoing electron).
This two-fold differential cross section can be written as a function of
transverse, longitudinal and interference parts $d^2\sigma_{\mbox{\tiny T}}$,
$d^2\sigma_{\mbox{\tiny L}}$, $d^2\sigma_{\mbox{\tiny TL}}$ and
$d^2\sigma_{\mbox{\tiny TT}}$, that only depend on $W$, $Q^2$ and $\theta^*$~:
\begin{eqnarray}
\frac{d^2\sigma}{d\Omega_{\pi}^\ast} &=&
\frac{d^2\sigma_{\mbox{\tiny T}}}{d\Omega_{\pi}^\ast}\ +
\ \epsilon\ \frac{d^2\sigma_{\mbox{\tiny L}}}{d\Omega_{\pi}^\ast}
+ \sqrt{2\epsilon(1+\epsilon)}\ \frac{d^2\sigma_{\mbox{\tiny TL}}}
{d\Omega_{\pi}^\ast}\cos{\phi} \nonumber \\
& & +\ \epsilon\ \frac{d^2\sigma_{\mbox{\tiny TT}}}{d\Omega_{\pi}^\ast}\cos{2\phi}.
\label{sigma_development}
\end{eqnarray}

In the rest of the paper, we will refer to the differential cross sections as
$\sigma_{\mbox{\tiny T}}+\epsilon\ \sigma_{\mbox{\tiny L}}$, $\sigma_{\mbox{\tiny TL}}$
and $\sigma_{\mbox{\tiny TT}}$.

%%%%% SECTION II %%%%%%%%%%%%%%%%%%%%%%%%%%%%%%%%%%%%%%%%%%%%%%%%%%%%%%%%%%%%%%%%
\section{Experimental Procedure and Data Analysis}
\label{sec:procedure}
%%%%%%%%%%%%%%%%%%%%%%%%%%%%%%%%%%%%%%%%%%%%%%%%%%%%%%%%%%%%%%%%%%%%%%%%%%%%%%%%%

%%%%%%%%%%%%%%%%%%%%%%%%
\subsection{Apparatus}
\label{subsec:apparatus}
%%%%%%%%%%%%%%%%%%%%%%%%

\begin{figure}
\includegraphics[width=8.6cm]{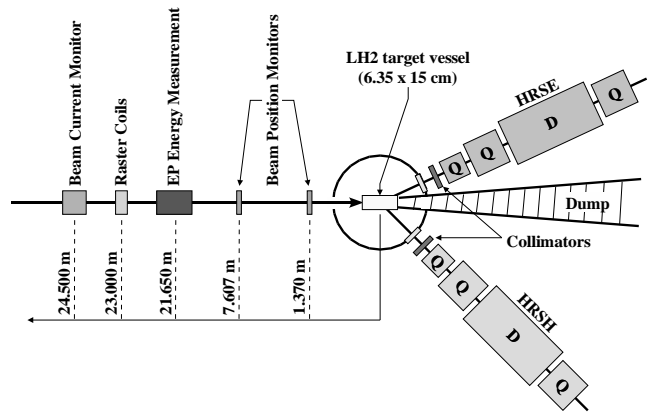}
\caption{\label{HALL-A}Layout of the Hall~A experimental setup.}
\end{figure}

The experiment was performed using a continuous electron beam with
an energy of 4032~MeV incident on a liquid hydrogen target.
The scattered electron and the recoil proton were detected in coincidence
in two high-resolution spectrometers (HRSE and HRSH).
Figure~\ref{HALL-A} is a top view of the experimental set-up and
the relevant components. More information on the Hall~A setup is
available in~\cite{Alcorn:2003}.

\subsubsection{Electron beam}
%%%%%%%%%%%%%%%%%%%%%%%%%%%%%

Typical beam intensities ranged from 60 to 120~$\mu$A; they were continuously
monitored during data taking using two resonant-cavity beam-current monitors
(BCM)~\cite{Denard:2001zg}. An absolute calibration of the BCMs was
performed at least once per day by employing an Unser transformer~\cite{Unser:1991dr}.
The measured standard deviation and drift of the BCM ensured a stability of
the current measurement of $\pm 0.3~\%$ over the entire experiment.
In order to avoid local boiling of hydrogen in the target, the incident beam
was rastered ($\pm 4$ mm horizontal and vertical)
with two asynchronized horizontal and vertical magnetic
coils ($\approx 20$~kHz) located 23~m upstream of the target.
The instantaneous position of the beam at the target was determined with an
accuracy of about 100~$\mu m$ with a pair of beam position monitors (BPM)
located at 7.607~m and 1.370~m upstream of the target~\cite{Hyde:2001}.
Each BPM is a resonant cavity with a set of four antenna wires parallel to the
beam axis. The difference between the signals on opposing wires is proportional to
the beam position.
%The raster amplitude was calibrated by measuring the spot
%size by moving a thin tungsten wire (harp scanner) through the beam.

\subsubsection{Target}
%%%%%%%%%%%%%%%%%%%%%%

The liquid hydrogen target material was contained in a cylindrical
aluminium vessel (0.0635~m diameter and 0.15~m long along the beam axis~---
see Fig.~\ref{HALL-A}). The target wall thickness was 175~$\mu$m. The
entrance and exit windows were 71 and 94~$\mu$m thick, respectively~\cite{Suleiman:1998}.

The target itself is located inside a cylindrical aluminum scattering
chamber connected to the beamline vacuum. The scattering chamber was equipped
with two 400~$\mu$m aluminium exit windows, each facing a spectrometer.

The working temperature and pressure of the hydrogen loop (19.0 K and 25 psia) %=1.72 bar
give a nominal density $\rho_0$ of $0.0723\ {\rm g/cm^3}$.

The data taken within 100~s after a substantial beam intensity variation
(e.g. beam trips) were excluded from the analysis to avoid instabilities in
the target density.

\subsubsection{Magnetic spectrometers}
%%%%%%%%%%%%%%%%%%%%%%%%%%%%%%%%%%%%%%

The two high resolution spectrometers (HRS) of QQDQ type are of identical
conception. Their main characteristics include a central momentum range from
0.3 to 4.0~GeV/c, and a nominal acceptance of $\pm~4.5$\% in momentum,
$\pm~65$~mrad in vertical angle, $\pm~30$~mrad in horizontal angle,
and $\pm~5$~cm in target length (transverse to the spectrometer
axis). The magnetic dipole in each spectrometer deflects the particle
trajectories in the vertical plane by 45$^{\circ}$ onto a 2~m
long focal plane. The acceptance is defined in part by a tungsten
collimator positioned at 1.109~m and 1.100~m (respectively for the
electron and hadron arms) from the target, and by the apertures of the magnets.
The vacuum box of the spectrometer is closed by a 178~$\mu$m Kapton entrance
window and a 100~$\mu$m titanium exit window. In this experiment,
the spectrometers were positioned with an absolute angular accuracy of 0.5~mrad.

\subsubsection{Detectors}
%%%%%%%%%%%%%%%%%%%%%%%%%

\begin{figure}
\includegraphics[width=8.6cm]{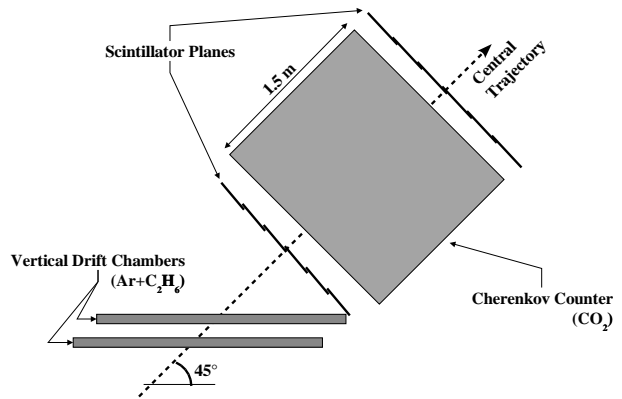}
\caption{\label{fig_detector}HRSE detector package. The vertical drift
chambers as well as the trigger scintillator hodoscopes are common to
both spectrometers.}
\end{figure}

The detector package of each spectrometer is shown in Fig.~\ref{fig_detector}.
It includes:
\begin{itemize}
   \item Two vertical drift chambers (VDC)~\cite{Fissum:2001st}, spaced by 50~cm, to
         define the trajectories of the charged particles; each VDC is equipped
         with two wire planes, to measure the intercepts and slopes of each
         trajectory in two perpendicular planes;
         charged particles passing within the acceptance of the spectrometer
         cross the plane of the chambers triggering from 3 to 5 sense wires.
         Each sense wire starts an updating time to digital converter (TDC) which
         is stopped by the acquisition trigger.

   \item Two scintillator planes S$_1$ and S$_2$ each consisting of 6 plastic
         scintillator paddles. The S$_1$ paddles are 29.3~cm (dispersive)
         by $36.0$~cm (transverse) and the S$_2$ paddles are $37.0$~cm
         (dispersive) by $60.0$~cm (transverse). In both planes the paddles
         overlap by 0.5~cm. Each paddle is viewed by 2 photomultiplier tubes
         (PMT) at opposite ends.

   \item A gas Cherenkov counter (filled with CO$_2$) viewed by 10 PMTs. Only
         the Cherenkov counter of the electron spectrometer was used in this
         experiment.
\end{itemize}
Each PMT output is fed to an amplitude to digital converter (charge integrating
ADC) and to a discriminator. Each discriminator signal is sent to a TDC and to
the fast electronics logic.

\subsubsection{Trigger electronics and data acquisition}
%%%%%%%%%%%%%%%%%%%%%%%%%%%%%%%%%%%%%%%%%%%%%%%%%%%%%%%%

The fast electronics logic defines several trigger signals for the data
acquisition system (DAQ) using the CEBAF Online Data Acquisition (CODA v1.4)~\cite{CODA}:
\begin{itemize}
\item T1 (T3) corresponds to a good electron (proton) event. It requires
      a coincidence between a paddle $i$ of the S$_1$ plane and a paddle
      $j$ of the S$_2$ plane within the directivity limits of the
      spectrometer ($|i-j|\le 1$). Each paddle event (S$_1$ or S$_2$)
      requires a coincidence between the two PMTs at the end of each paddle.

\item T2 (T4) defines a deficient electron (proton) event. This requires
      that either the S$_1$ $\cap$ S$_2$ coincidence is not within the directivity
      limits ($|i-j| > 1$) or that only one scintillator plane fires.
      For the T2 trigger, if only one scintillator plane has a two-ended coincidence,
      the trigger logic requires a coincidence with the gas Cherenkov counter signal.

\item T5 is the main trigger and is defined by a coincidence of T1 and T3 within 100~ns.
\end{itemize}

Although all triggers can fire the DAQ, T5 has priority while
other triggers are prescaled. This fraction is
set using prescale factors (PS$_1$, PS$_2$, PS$_3$ and PS$_4$). The
encoding of the analog signals and the transfer of the digitized signal
to the computer buffers takes $\sim$ 700~$\mu$s. When the DAQ is triggered,
it forbids any other trigger until the first is processed. This induces
acquisition dead times up to 30~\% for some high counting rate conditions.
The number of events for every trigger type is recorded by a running scaler,
which is read and logged by the DAQ every 10 seconds.

%%%%%%%%%%%%%%%%%%%%%%%%%%%%%%%%%
\subsection{Data taking geometry}
\label{subsec:data}
%%%%%%%%%%%%%%%%%%%%%%%%%%%%%%%%%

Data were taken at nine spectrometer angle and momentum settings (numbers
4-12 in Table~\ref{settings}), covering the entire resonance region, i.e.
a total CM energy $W$ varying between pion threshold and 2.0~GeV. $W$ is
the invariant mass of the $(\gamma^*p)$ system, $W=\sqrt{(k-k'+p)^2}$.
The acceptance in $\theta^*$ was centered around $180^\circ$.
Complementary measurements (settings \#1, 2, 3 in Table~\ref{settings})
were included in order to increase the statistical accuracy around the
pion production threshold.
Additional H$(e,e')p$ elastic scattering measurements with a sieve
slit (and both spectrometers tuned to electrons) and
Al,~C$(e,e')X$ quasi-elastic measurements with an array of foil targets
served for calibration of detectors and spectrometer optics.
The relevant information on production data is summarized in Table~\ref{settings}.

\begingroup
\squeezetable
\begin{table}[ht]
\caption{\label{settings}Kinematical settings used for the $\pi^0$ data taking.
         The incident electron energy was 4032~MeV (discussed in
         Section~\ref{subsub:method}). The values shown in the table are the central
         values within the acceptance.}
\begin{ruledtabular}
\begin{tabular}{cccccc}
Setting &$W_{\mathrm nom}$&$k'_{\mathrm nom}$&$(\theta_e)_{\mathrm nom}$&
$p'_{\mathrm nom}$&$(\theta_p)_{\mathrm nom}$ \\
number & (GeV) & (GeV/c) &($^\circ$)&(GeV/c)&($^\circ$) \\ \hline
 1 & 1.180 & 3.433 & 15.43 & 1.187 & -50.00 \\ %res\_1.2a
 2 & 1.178 & 3.433 & 15.43 & 1.187 & -48.50 \\ %res\_1.2b
 3 & 1.177 & 3.433 & 15.43 & 1.187 & -46.50 \\ %res\_1.2c
 4 & 1.217 & 3.282 & 15.77 & 1.323 & -45.41 \\ %res\_1.30
 5 & 1.252 & 3.176 & 16.04 & 1.418 & -41.67 \\ %res\_1.50
 6 & 1.326 & 3.043 & 16.39 & 1.539 & -37.49 \\ %res\_1.75
 7 & 1.431 & 2.909 & 16.76 & 1.662 & -33.82 \\ %res\_2.00
 8 & 1.526 & 2.776 & 17.16 & 1.787 & -30.60 \\ %res\_2.25
 9 & 1.613 & 2.642 & 17.87 & 1.914 & -27.75 \\ %res\_2.50
10 & 1.690 & 2.482 & 18.15 & 2.067 & -24.75 \\ %res\_2.80
11 & 1.795 & 2.269 & 18.99 & 2.274 & -21.34 \\ %res\_3.20
12 & 1.894 & 2.056 & 19.96 & 2.482 & -18.46 \\ %res\_3.60
\end{tabular}
\end{ruledtabular}
\end{table}
\endgroup

%%%%%%%%%%%%%%%%%%%%%%%%%%
\subsection{Data analysis}
\label{subsec:analysis}
%%%%%%%%%%%%%%%%%%%%%%%%%%

\subsubsection{Method}
%%%%%%%%%%%%%%%%%%%%%%
\label{subsub:method}

The data analysis procedure includes several passes. In a first step,
we reject any sequence of CODA events
collected when one of the stability requirements fails:
beam intensity or position, spectrometer magnetic elements, etc.

Next, the Hall~A analyzer ESPACE (Event Scanning Program for Hall~A
Collaboration Experiments)~\cite{Offerman:1997} is used to construct the trajectory
of the particles in the spectrometer focal plane from the VDC data~:
two position coordinates $X_{\rm fp}$ and $Y_{\rm fp}$ and two cartesian
angles $\phi_{\rm fp}$ and $\theta_{\rm fp}$.

Then, using the beam position information at the target and the database
for the spectrometers optics, ESPACE reconstructs the entire kinematics
of the electron and the proton at the vertex, as well as the interaction point.
This database has been optimized for the kinematical settings of this
experiment~\cite{Jaminion:2001th}.
Both particles at the vertex are described with four spectrometer variables~:
the transverse coordinate $Y_{\rm tg}$, the two cartesian angles $\phi_{\rm tg}$
and $\theta_{\rm tg}$, and the relative momentum
\begin{eqnarray}
\delta k' &=& \frac{k'-k'_{\rm nom}}{k'_{\rm nom}}\; \mathrm{(electron)}\\
\delta p' &=& \frac{p'-p'_{\rm nom}}{p'_{\rm nom}}\; \mathrm{(proton)}.
\end{eqnarray}
The dispersive coordinate $X_{\rm tg}$ is deduced from the beam
information. The energy loss in the target and spectrometer windows is
also taken into account.

At this stage, the position and the shape of the missing mass squared
$M_X^2$ distribution are indicators of how the positioning
of the spectrometers and the beam are under control.
The missing mass resolution was optimized by varying the beam energy,
the vertical angle $\theta_{\rm tg}$ of the electron arm, the horizontal
angles of both arms $\phi_{\rm tg}$, and the calibration of the vertical
beam raster amplitude at the target. The result of this optimization yields
an average correction for the beam energy of $-13$~MeV to the nominal value
of 4045~MeV, with a dispersion of $\pm$~3~MeV, varying from one run to another.
A similar procedure based on the horizontal position of the reconstructed
vertex is used to determine the calibration of the horizontal raster amplitude
and the horizontal mispointing of the spectrometers.

\subsubsection{Simulation and radiative corrections}
\label{subsubsec:simulation}
%%%%%%%%%%%%%%%%%%%%%%%%%%%%%%%%%%%%%%%%%%%%%%%%%%%%

We use a detailed simulation~\cite{VanHoorebeke} which takes into account all
processes that affect the characteristics of the experimental
data. Indeed, these data stem from a convolution of the ``ideal'' events
defined at the vertex with a number of processes that influence the
incident beam and the outgoing (detected) particles.
The simulation incorporates the beam profile distribution, collisional
energy losses, multiple scattering, internal and external bremsstrahlung and
radiative corrections~\cite{Vanderhaeghen:2000ws}, as well as other
resolution effects ({\it e.g.\/} from optics and detector resolution).
The spectrometers acceptance is simulated with a model based on the optical
design of the spectrometer and field maps of the magnets~\cite{Jaminion:2001th}.
In order to reconcile the results of the simulation with the data,
an additional smearing had to be introduced. This correction depends
on the data-taking geometry and is listed in Table~\ref{resolution}.

Events are generated according to a model cross section $d\sigma_{\rm model}$.
In a first step, $d\sigma_{\rm model}$=MAID2000~\cite{Drechsel:1998hk,Kamalov:2001yi}
(this is discussed in Section~\ref{sec:wphidep}). In a second step a {\em local fit} based on
MAID2003~\cite{Tiator:2003uu} is performed on our data (Section~\ref{subsec:MAID}),
and in a third step a dependence on $Q^2$ is added based on our experimental results
(Section~\ref{sec:Q2dep}).

\begingroup
\squeezetable
\begin{table}[ht]
\caption{\label{resolution}Additional Gaussian resolution smearing at each
         experimental setting (rms) for reconstructed variables at the target.}
\begin{ruledtabular}
\begin{tabular}{ccccccc}
&\multicolumn{2}{c}{electron} &\multicolumn{4}{c}{proton} \\ \hline
Setting & $\theta_{\rm tg}$ & $Y_{\rm tg}$ &
$\delta p'$ & $\phi_{\rm tg}$ & $\theta_{\rm tg}$ & $Y_{\rm tg}$ \\
number & (mrad)& (mm)&(10$^{-4}$)&(mrad)&(mrad)&(mm) \\ \hline
 1 & 1.00 & .00 & 2.00 & 1.00 & 2.00 & .30 \\ %res\_1.2a
 2 & 1.00 & .00 & 2.00 & 1.00 & 2.00 & .30 \\ %res\_1.2b
 3 & 1.00 & .00 & 2.00 & 1.00 & 2.00 & .30 \\ %res\_1.2c
 4 & 1.35 & .30 & 2.75 & 1.35 & 2.75 & .30 \\ %res\_1.30
 5 & 1.45 & .42 & 3.00 & 1.45 & 3.00 & .42 \\ %res\_1.50
 6 & 1.80 & .33 & 3.60 & 1.80 & 3.60 & .33 \\ %res\_1.75
 7 & 1.65 & .00 & 3.30 & 1.65 & 3.30 & .00 \\ %res\_2.00
 8 & 1.80 & .66 & 3.60 & 1.80 & 3.60 & .66 \\ %res\_2.25
 9 & 1.80 & .66 & 3.60 & 1.80 & 3.60 & .66 \\ %res\_2.50
10 & 1.80 & .66 & 3.60 & 1.80 & 3.60 & .66 \\ %res\_2.80
11 & 1.80 & .66 & 3.60 & 1.80 & 3.60 & .66 \\ %res\_3.20
12 & 1.50 & .66 & 3.00 & 1.50 & 3.00 & .66 \\ %res\_3.60
\end{tabular}
\end{ruledtabular}
\end{table}
\endgroup

Our procedure for radiative corrections has been actually developed
for process (\ref{react2}) following the exponentiation method of
\cite{Vanderhaeghen:2000ws}, and has been applied in the same way to
process (\ref{react1}). In this method, radiative corrections are
implemented in two parts according to the source of photon radiation.
The first contribution is the acceptance-dependent part of the internal
and external bremsstrahlung from the electron lines, and as such it is
included in the simulation~\cite{VanHoorebeke}.
This reproduces the radiative tail in the missing mass squared spectrum
(see Fig.~\ref{fig-mx2}). The second contribution is expressed as
a constant factor equal to $0.93$ at $Q^2=1.0$~GeV$^2$
applied to the cross section. The systematic error associated with the
radiative corrections is taken equal to $\pm$2\%~\cite{Vanderhaeghen:2000ws}.
%Recently a new procedure has been proposed to calculate radiative
%corrections to exclusive $\pi^0$ electroproduction~\cite{Afanasev:2002ee}.
%Unfortunately, application of this method to our data would require
%a completely different analysis, which is beyond the scope of this work.

\subsubsection{$\pi^0$ event selection}
\label{subsubsec:cuts}
%%%%%%%%%%%%%%%%%%%%%%%%%%%%%%%%%%%%%%%

\begin{figure}
\includegraphics[width=8.6cm]{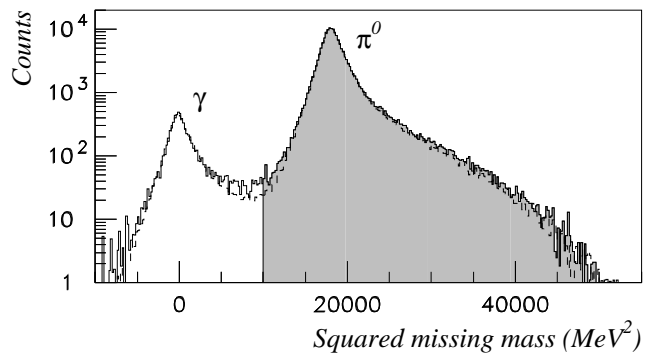}
\caption{\label{fig-mx2}Experimental distribution of the missing mass
         squared (solid) and the corresponding simulated spectrum
         (dashed line) obtained by registering the process p(e,e'p)X.
         The peak around zero corresponds to events originating from
         reaction~(\ref{react2}). The maximum of the second peak is at
         the pion mass squared $m_{\pi0}^2 = 18.2\cdot 10^3$~MeV$^2$
         as to be expected if events originated from reaction~(\ref{react1}).
         The colored region represents the event distribution within
         the selection window for process~(\ref{react1}).}
\end{figure}

The following criteria and cuts have been applied to properly select the
$\pi^0$ events~:
\begin{itemize}
\item A suitable coincidence timing window, as illustrated in Fig.~\ref{fig-tccor}.

\begin{figure}
\includegraphics[width=8.6cm]{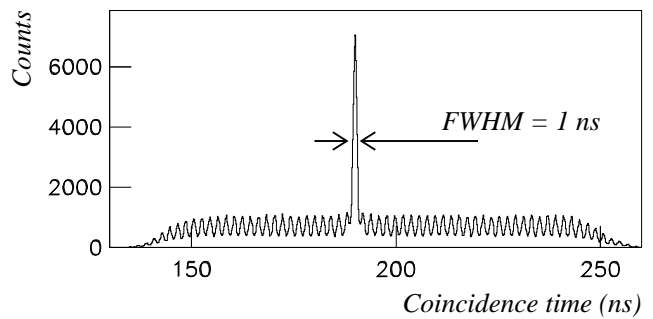}
\caption{\label{fig-tccor}Coincidence time for setting \#7. The time is corrected
         for the path length in the spectrometer and for the proton velocity.
         The fine structure in the time spectrum is due to the 500~MHz structure
         of the beam.}
\end{figure}

\item A directivity cut applied on the particle's position in the
      collimator plane at the entrance of each spectrometer~:
      $\pm 2.9$~cm in horizontal and $\pm 5.8$~cm in vertical (This corresponds
      to 87\% of the total geometrical acceptance of the 6~msr collimator).
\item An acceptance cut defined for both arms by~:
\begin{equation}
\delta k'(p') \le A\pm B*\phi_{\rm tg}+C*Y_{\rm tg}^2
\end{equation}
      with $A\ =\ 0.17$, $B\ =\ 6.0$~rad$^{-1}$ and $C\ =\ -23.15$~m$^{-2}$.
      This cut approximates the dipole aperture and was used to symmetrise
      the acceptance which is not completely defined by the collimator alone.
\item An acceptance cut defined in both arms in the plane ($Y_{\rm tg}$,
      $\phi_{\rm tg}$). This cut has an hexagonal shape and tends to
      reproduce the quadrupoles apertures. More information on this cut is
      given in \cite{Laveissiere:2001th}.
\item A cut on the horizontal transverse distance $d$ between the beam
      and the reconstructed vertex (using both arms)~:
\begin{equation}
\left| d \right|\  <\ 0.003~\mathrm{m}
\end{equation}
\item A selection window on the missing mass squared~:
      $10000\ <\ M_{x}^{2} <\ 50000$~MeV$^2$.
      The lower boundary of the selection window serves to suppress
      the yield from reaction~(\ref{react2}) which is manifest as
      a peak at $M_X^2=0$ in Fig.~\ref{fig-mx2}.
\end{itemize}

%%%%%%%%%%%%%%%%%%%%%%%%%%%%%%%%%%%%%%
\subsection{Cross sections evaluation}
\label{subsec:cross}
%%%%%%%%%%%%%%%%%%%%%%%%%%%%%%%%%%%%%%

\subsubsection{Extraction method}
\label{subsubsec:method}
%%%%%%%%%%%%%%%%%%%%%%%%%%%%%%%%%

In the present analysis, a typical experimental bin of phase space is defined in
the five kinematic variables $Q^2, W, \epsilon$, $\cos{\theta^*}$, $\phi$, and the
missing mass squared $M_X^2$.
The number of events in each  bin is the product of the integrated luminosity
${\cal L}$ and the convolution of the physical cross section with resolution
effects over all the experimental acceptance.
Let $N_i$ denote the number of counts observed in bin $i$,
and ${\cal K}_i$ the experimental resolution and acceptance function
of the same bin. Then
\begin{equation}
N_i={\cal L}\,\int{\left[d^6\sigma\otimes{\cal K}_i\right]}.
\end{equation}

The number of events simulated in the given bin $i$,
$N_i^{(s)}$, is defined by the simulated luminosity ${\cal L}_s$ times
the five-fold differential Born (non-radiative) cross section $d^5\sigma_s$
depending on variables $Q^2$, $W$, $\epsilon$, $\cos{\theta^*}$ and $\phi$
convoluted with the radiative process and the experimental resolution.
If we denote the contribution of the radiative processes (including internal
and external bremsstrahlung)  by $d{\cal R}_s$, and the simulated resolution
response and acceptance of bin $i$  by ${\cal K}_i^{(s)}$, the simulated
number of events is then~:
\begin{equation}
N_i^{(s)}={\cal L}_s\,\int{\left[
d^5\sigma_{\rm model}\otimes d{\cal R}_s\otimes{\cal K}_i^{(s)}\right]}.
\label{convolution}
\end{equation}
If the processes described by Eq.~(\ref{convolution}) are correctly
taken into account in the simulation, then,
assuming that the relative variation of the true cross section
and the simulated one around a point
${\cal P}_0=(W,Q^2,\epsilon,\cos{\theta^*},\phi)$ are the same~:
\begin{equation}
\frac{ d\sigma({\cal P})-d\sigma({\cal P}_0)}{d\sigma({\cal P}_0)} =
\frac{ d\sigma_{\rm model}({\cal P})-d\sigma_{\rm model}({\cal P}_0)}
{d\sigma_{\rm model}({\cal P}_0)}~,
\label{condition}
\end{equation}
we arrive at the experimental differential cross section at point
${\cal P}_0$~:
\begin{equation}
d^5\sigma({\cal P}_0)=\frac{{\cal L}_s}{{\cal L}}
\times\frac{N_i}{N_i^{(s)}}\times
d^5\sigma_{\rm model}({\cal P}_0).
\label{sig-exp}
\end{equation}
These assumptions are verified {\it a posteriori} by observing a good
agreement between the experimental and simulated distributions
(e.g. missing mass spectra in Fig.~\ref{fig-mx2}).
The size of the bins is only constrained by the magnitude of the
resolution and radiative effects, and by the variation of the model
cross section $d\sigma_{\rm model}$.
In this analysis, we choose the point ${\cal P}_0$ to lie at the
center of each bin.

%%%%%%%%%%%%%%%%%%%%%%%%%%%%%%%%%%%%%%%%%%%%%%%%%%%%%%%%%%%%
\subsubsection{Adjustment of the model parameters}
%%%%%%%%%%%%%%%%%%%%%%%%%%%%%%%%%%%%%%%%%%%%%%%%%%%%%%%%%%%%

The procedure described by Eq.~\ref{condition} to evaluate the
experimental cross section relies heavily on the accuracy of
the simulation of the true cross section inside each bin
by the model cross section $d\sigma_{\rm model}$. Thus, it is
imperative to employ a realistic model cross section in the
Monte Carlo simulation.

At the start of the analysis we used in the simulation and,
consequently, in the determination of the experimental cross section,
the MAID2000 model (see Section~\ref{subsec:raw}).
It was found at this stage of the analysis that the model cross section
departs from the measured one, especially for the second and third resonance
regions. In particular we observed strong differences in the $W$,
$Q^2$ and $\phi$ dependences of the cross section which motivated an
adjustment of the model parameters (see Section~\ref{subsec:MAID}).

At the second step of the iteration the experimental cross section
was evaluated by employing the model version MAID2003
with adjusted parameters ({\em local fit}). This adjustment did not involve
model parameters responsible for the $Q^2$ dependence
because the present data comprise a rather limited $Q^2$ interval.

In a last step we used our experimental results to obtain an estimation
of the $Q^2$ dependence. Another iteration was performed afterwards by
including in the simulation the new $Q^2$ dependence (see Section~\ref{sec:Q2dep}).
The final results are presented in Sections~\ref{sec:Q2dep} and~\ref{sec:results}.

\subsubsection{Corrections and systematic errors}
%%%%%%%%%%%%%%%%%%%%%%%%%%%%%%%%%%%%%%%%%%%%%%%%%
\label{subsub:errors}

In the extraction of the cross section values, a number of corrections
have to be taken into account. For each correction, the residual
systematic error was evaluated. All relevant quantities are given
in Table~\ref{systematic}.

\begingroup
\squeezetable
\begin{table}[ht]
\caption{\label{systematic}Correction and systematic error evaluation}
\begin{ruledtabular}
\begin{tabular}{lrr}
Source                        &    Correction  &  Induced error on $\sigma$\\ \hline
Trigger efficiency            &        1-10~\% &    $\pm$0.0~\% \\
Acquisition dead time         &        0-30~\% &    $\pm$0.0~\% \\
Electronics dead time         &     2.5-4.5~\% &    $\pm$0.1~\% \\
Tracking efficiency           &     3.0-8.0~\% &    $\pm$0.5~\% \\
Optics                        &                &    $\pm$1.2~\% \\
Acceptance                    &                &    $\pm$2.0~\% \\
Target boiling                &                &    $\pm$1.0~\% \\
Proton absorption correction  &         1-3~\% &    $\pm$0.1~\% \\
Radiative corrections         &                &    $\pm$2.0~\% \\
Photon contamination          &         1.0~\% &    $\pm$0.0~\% \\ \hline
Quadratic sum                 &                &    $\pm$3.3~\% \\
\end{tabular}
\end{ruledtabular}
\end{table}
\endgroup

The trigger efficiency correction $E_{1,2}(x,y)$ is calculated run by run
for each scintillator plane (1 and 2), locally in longitudinal $(x)$ and
transverse $(y)$ directions. The VDCs determine the particle's track
location in the scintillator. For this efficiency study, a stringent event
selection in the four planes of the VDCs is applied. The efficiency correction
factor is then (for the electron arm)~:
\begin{eqnarray}
\lefteqn{E_{1,2}(x,y)\ =\ 1\ +} \\
\nonumber
& & \frac{N(T_2.\overline{S_{1,2}})\times PS_2}
{N(T_5)+N(T_1)\times PS_1+N(T_2.\overline{S_{1,2}})\times PS_2 }~.
\end{eqnarray}
Here, $N(T_2.\overline{S_{1,2}})$ is the number of T2 trigger events
tracked by the VDCs to the $(X_{fp},Y_{fp})$ area, but with no $S_1~\cap~S_2$
scintillator coincidence.
A similar procedure is applied for the hadron arm with the T4 triggers.
The correction is of the order of 2~\% for the electron arm, and less than 1\% for
the hadron arm. The accuracy on this correction is governed by the number of
T2 and T4 triggers, and is of the order of 1--5\% of the inefficiency.
This induces no appreciable systematic error in the final result.

The dead-time correction factor of the acquisition system is the ratio
of the number of events measured by the scaler associated to trigger T5
to the total number of coincidence events found in each run. It ranges
from 0 to 30~\% and the associated error is negligible.

The dead time associated with the electronics is defined by the setup and
depends directly on the beam intensity. It is evaluated for each run from
the singles rate of each discriminator associated with the scintillator
paddles and electron Cherenkov. Typically, the correction is 2.5\% at the
$\Delta(1232)$ resonance, and 4.5\% for the highest $W$ setup. The induced
systematic error is negligible.

The intrinsic efficiency of the VDCs is determined by the efficiency of each
sense wire. A good track requires a signal from at least 3 wires in each plane.
The fact that a typical track intercepts five cells in each wire plane makes
the VDCs global inefficiency negligible.

The tracking efficiency is affected by accidental hits, caused by background
events, which can prevent the algorithm from reconstructing the good track.
Thus, a noticeable fraction of the events has more than one reconstructed track
in the VDCs. These events are rejected in the analysis and the luminosity is
decreased in proportion. Also this correction depends strongly on the setup
configuration and on the beam current, and varies between 3\% and 8\%.
The systematic error in this correction is estimated to be 10\% of the correction.

Independent of the uncertainty in the total acceptance of the HRS pair,
we have a cross section uncertainty from the imperfect knowledge of the
spectrometer optics.
We subdivide the  acceptance into bins in the physics variables
$Q^2$, $W$, $M_X^2$, $\cos{\theta^*}$, and $\phi$.
The precise volume of each bin is subject to uncertainties due to {\it local\/}
variations in the average reconstruction of vertex variables. We estimate these
uncertainties from the rms deviations between the positions of the sieve slit
holes (at the entrance of each spectrometer) and the mean reconstructed position
of these holes.
Local variations in the calibration of vertex positions along the beam line
influence the luminosity, which is proportional to the effective target length
viewed by the HRS pair.
We estimate the uncertainty in the effective target lengths from the deviations
between the positions of a set of seven reference foil targets
and their reconstructed positions.
The sieve slit holes are on a square grid of spacing 25~mm vertical and 12.5~mm
horizontal. The seven targets were located at 0, $\pm 20$, $\pm 50$, and $\pm 75$~mm
along the beam axis. In the electron arm, the rms deviations of the mean
reconstructed values are 0.065~mm  and 0.050~mm, for the vertical and horizontal sieve
slit holes, respectively, and 0.145~mm for the target foils along the beam axis. The
same values for the hadron arm are 0.097~mm, 0.027~mm, and 0.220~mm.
Dividing the rms variations by the respective spacings in the vertical and
horizontal sieve slit holes and the reference targets, we obtain the
contributions to the cross section uncertainties arising from local
variations in optics. Adding all contributions in quadrature yields an
uncertainty of $\pm$1.2\%. This is the optics uncertainty in Table~\ref{systematic}.

We have performed a set of acceptance cuts to improve the
agreement between the experiment and the simulation (see
Section~\ref{subsubsec:cuts}). The uncertainty associated
with possible residual discrepancies is estimated to be $\pm 2$\%.

The beam current and its variation can lead to target density
corrections to the luminosity.  The beam was rastered over an
area proportional to the beam current, and equal to $8\times 8$~mm$^2$
at $100\ \mu$A to minimize any effect of hydrogen boiling.
In analyses of single-arm elastic data, no correlation was observed
between the target density and the beam current~\cite{Jutier:2001th}, within $\pm 1\%$.

From the analysis of the single-arm elastic data, it was also concluded that
the target impurity is negligible, i.e. $\leq\ 0.02$\%.

A correction was also evaluated for lost recoil protons, either from
interactions with the liquid hydrogen target material or in the different
windows; its value equals 1\% near the pion production threshold and
reaches about 3\% at the highest $W$.

The error associated with the radiative correction is $\pm 2$\%. This
matter has been discussed in Section~\ref{subsubsec:simulation}.

Finally, at low $W$, the contribution of reaction~(\ref{react2}) is not
negligible in the selected window in missing mass squared. A correction
has been made to subtract the photon events located below the $\pi^0$
peak (see Fig.~\ref{fig-mx2})~; it does not induce any further systematic
error.

The total error evaluated as a quadratic sum of all the contributions
amounts to $\pm$3.3~\%. This total error will be added quadratically to the
model dependence error discussed in Section~\ref{sec:results}.

%%%%% SECTION III %%%%%%%%%%%%%%%%%%%%%%%%%%%%%%%%%%%%%%%%%%%%%%%%%%%%%%%%%%%%%%%
\section{Study of the $W$, $\theta^*$ and $\phi$ dependences}
\label{sec:wphidep}
%%%%%%%%%%%%%%%%%%%%%%%%%%%%%%%%%%%%%%%%%%%%%%%%%%%%%%%%%%%%%%%%%%%%%%%%%%%%%%%%%

%%%%%%%%%%%%%%%%%%%%%%%%%%%%%%%%%%%%%%%%%%%%%%%%%%%%%%%%
\subsection{Excitation curves and angular distributions}
\label{subsec:raw}
%%%%%%%%%%%%%%%%%%%%%%%%%%%%%%%%%%%%%%%%%%%%%%%%%%%%%%%%

\begin{figure}
\includegraphics[width=8.6cm]{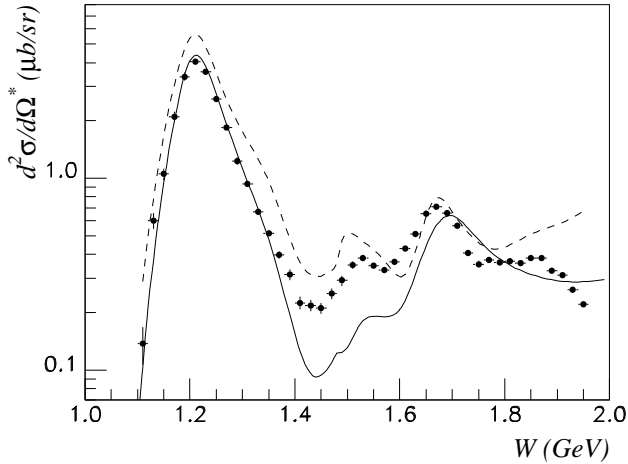}
\caption{\label{fig-wscan}Excitation curve for $\gamma^* p \to p \pi^0$
         at $Q^2$ = 1.0~GeV$^2$, $\cos{\theta^*}$ = --0.975, and $\phi = 75^\circ$. The full
         line corresponds to MAID2000~\cite{Drechsel:1998hk,Kamalov:2001yi} and the dashed line
         to SAID ({\em NF18K solution})~\cite{Arndt:1996ak}.}
\end{figure}

The method presented in Section~\ref{subsec:cross} was applied to produce the two-fold
differential cross section as a function of $W$, $Q^2$, $\cos{\theta^*}$ and $\phi$.
The cross section is evaluated in $50\times1\times4\times12$ kinematical
intervals chosen as shown in Table~\ref{binning}. Figure~\ref{fig-wscan} shows
a sample excitation curve for $Q^2=1.0$~GeV$^2$, $\cos{\theta^*}=-0.975$
and $\phi =75^\circ$.

\begingroup
\squeezetable
\begin{table}[ht]
\caption{\label{binning}Binning intervals for each variable. Note that the
analysis of the $Q^2$ dependence discussed in Section~\ref{sec:Q2dep}
required splitting of the $Q^2$ range into 6 intervals as well as a wider
binning in $W$ and $\cos{\theta^*}$ (these values are indicated in parenthesis).}
 \begin{ruledtabular}
\begin{tabular}{llll}
Variable         & ~~~~~ Range            & Number of & ~~~~ Interval       \\
                 &                        & intervals & ~~~~~ width         \\ \hline
$W$              &  [1.00;2.00]~GeV       & 50(10)    & 0.02(0.1)~GeV       \\
$Q^2$            &  [0.85;1.15]~GeV$^2$   &   1(6)    & 0.3(0.05)~GeV$^2$   \\
$\cos{\theta^*}$ & ~~ [-1;-0.8]           &   4(1)    & 0.05(0.2)           \\
$\phi$           & ~ ~  [0;360]~$^\circ$  &     12    & ~~ ~~ ~ 30~$^\circ$ \\
\end{tabular}
\end{ruledtabular}
\end{table}
\endgroup

\begin{figure}
\includegraphics[width=8.6cm]{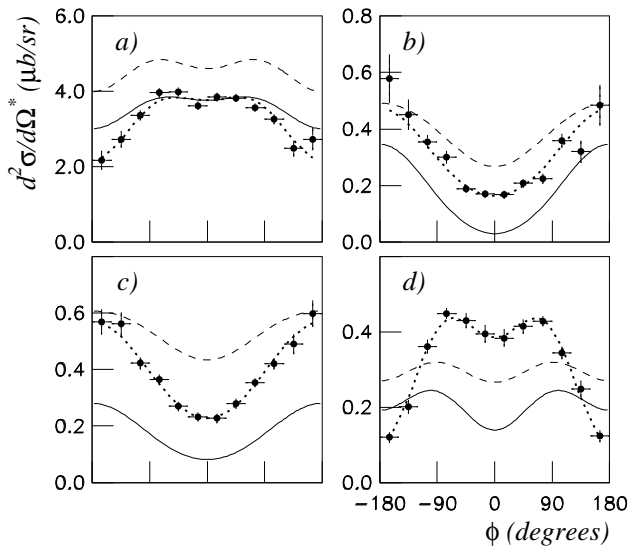}
\caption{\label{fig-phiscan}Azimuthal
         angular distributions for $\gamma^* p \to p \pi^0$ at $Q^2$ = 1.0~GeV$^2$,
         $\cos{\theta^*}$ = --0.975 for different points in $W$:
         1230~MeV --- a), 1410~MeV --- b), 1510~MeV --- c) and 1610~MeV --- d).
         The solid curve corresponds to MAID2000~\cite{Drechsel:1998hk,Kamalov:2001yi}
         and the dashed line to SAID ({\em NF18K solution})~\cite{Arndt:1996ak}.
         The dotted line approximating the data points is obtained from the fit of
         Eq.~(\ref{sigma_development}).}
\end{figure}

The data integrated over the whole $Q^2$ range yield the cross section
as a function of $\phi$ for each $W$ and $\cos{\theta^*}$ interval.
As an example, we present  in Fig.~\ref{fig-phiscan} the azimuthal
distributions for four points in $W$.

The corresponding cross-section data evaluated in the framework of the
MAID2000 model~\cite{Drechsel:1998hk,Kamalov:2001yi} demonstrate a good agreement with the
results obtained in the $\Delta(1232)$ region. The agreement
deteriorates as $W$ increases (Figs.~\ref{fig-wscan} and \ref{fig-phiscan}).

For each bin in $W$ and $\cos{\theta^*}$, we obtain the separated
cross sections $\sigma_{\mbox{\tiny T}}+\epsilon\ \sigma_{\mbox{\tiny L}}$,
$\sigma_{\mbox{\tiny TL}}$ and $\sigma_{\mbox{\tiny TT}}$ by fitting
Eq.~(\ref{sigma_development})
%% a second order polynomial function in $\cos\phi$ 
to the data in 12 bins in $\phi$ (Fig.~\ref{fig-phiscan}).
In the procedure of minimization only the statistical errors are used.

There exists another well-developed technique for describing
pion electroproduction over the whole resonance region --- the SAID
analysis~\cite{Arndt:1996ak,Arndt:2002xv,Arndt:2003zk}. SAID employs the regularly
updated compilation of available data for photo- and electro-production
reactions to constrain a certain set of parameters in an energy-dependent
multipolar fit. The output of such fit corresponding to the {\em NF18K solution}
is displayed in Fig.~\ref{fig-wscan}.
Although the {\em NF18K solution} overpredicts our data from threshold to $W=1.6$~GeV,
the general trends of the cross section are well reproduced (Figs.~\ref{fig-wscan},
\ref{fig-phiscan}). The agreement between the SAID model and our data is
significantly improved after the data is added to the world database (see
Section~\ref{subsec:SAID}).

%%%%%%%%%%%%%%%%%%%%%%%%%%%%%%%%%%%%%%%%%%%%%
\subsection{Amplitude Analysis with MAID}
\label{subsec:MAID}
%%%%%%%%%%%%%%%%%%%%%%%%%%%%%%%%%%%%%%%%%%%%%

With our complete dataset of 363 data points in three observables and
three values of pion emission angle we performed a data analysis using
the unitary isobar model MAID2000~\cite{Drechsel:1998hk,Kamalov:2001yi}.
This model is based on the evaluation of a nonresonant background described
by Born terms and vector meson exchange, and a resonant part modeled with
Breit-Wigner functions for all four star nucleon resonances below $W=2$~GeV,
\begin{equation}
t_{\gamma\pi}^{\alpha}=v_{\gamma\pi}^{bg,\alpha}(1+it_{\pi}^{\alpha})+t_{\gamma\pi}^{BW,\alpha}
e^{i\phi_\alpha}\,.
\end{equation}
Both parts are individually unitarized. For the background part this is
done in the usual K-matrix approximation and for the resonance part by
including an energy dependent unitarization phase $\phi_\alpha$. The
background and the hadronic parameters of the resonances are fixed,
leaving only the electromagnetic couplings of the $N^*$'s and $\Delta$'s
as free parameters. For electroproduction these are electric, magnetic
and longitudinal couplings that can be expressed in terms of the helicity
amplitudes $A_{1/2}$, $A_{3/2}$ and $S_{1/2}$. They are defined at the
resonance position $W=M_R$ and are related to the transition form factors.

In MAID2000 the $Q^2$ dependence of these couplings is modeled by
semi-phenomenological form factors. In the MAID2003 calculation~\cite{Tiator:2003uu},
it has a phenomenological form fitted to all existing electroproduction data
({\em global fit}). Since our data are taken in a narrow interval around
$Q^2=1.0$~GeV$^2$, the current analysis will be a fixed-$Q^2$ analysis
({\em local fit}).

In our  analysis, for the 13 nucleon resonances below
$W=2$~GeV, we fix the parameters of five from the results of the {\em global fit},
and adjust the parameters of the remaining eight resonances.  These
are the $P_{33}(1232)$, $P_{11}(1440)$, $D_{13}(1520)$, $S_{11}(1535)$,
$S_{31}(1620)$, $S_{11}(1650)$, $F_{15}(1680)$ and $D_{33}(1700)$,
giving a total of 20 free parameters. 
In order to estimate the model uncertainties in our fit, we successively
fixed individual resonance parameters to the values of the {\em global fit},
and investigated the fluctuations in the remaining parameters.
In Table~\ref{resonances}
we give the result of our {\em local fit} for the five resonances for which the
parameters are found to reasonably fluctuate around inital values.
The $S_{31}(1620)$, $S_{11}(1650)$, and $D_{33}(1700)$ are excluded from
the table, as their parameters could not be constrained in the present
analysis.
The main reason for this is a restricted angular range of our dataset, which
is confined to backward angles.  Furthermore, by not including any world
$\pi^+$ data, our fit is insensitive to isospin.  Even so, our data show
strong sensitivity to the resonances at large $W$.
In Table~\ref{resonances}, we compare our MAID2003~\cite{Tiator:2003uu}
{\em local fit} with the default values of MAID2000.
For the $\Delta(1232)$ resonance, we give in addition the
$R_{EM} = E2/M1$ and the $R_{SM} = C2/M1$ ratios.
Both MAID2003 values are consistent with the previous MAID2000
fits~\cite{Drechsel:1998hk,Kamalov:2001yi}. 
The $R_{SM}$ ratio is very well determined by
$\sigma_{\mbox{\tiny TL}}$ and tends to larger negative values at
$Q^2=1.0$~GeV$^2$ in comparison to $Q^2=0$, while the $R_{EM}$ ratio
is loosely constrained.
% is much more uncertain and also the model uncertainties are larger than
Furthermore, the model uncertainties are larger for latter value than
for the $C2/M1$ ratio. From $\sigma_{\mbox{\tiny TL}}$ we also find a
large sensitivity to the $S_{0+}$ amplitude of the $S_{11}(1535)$
resonance in the minimum around $W=1500$~MeV.

\begingroup
\squeezetable
\begin{table}[ht]
\caption{\label{resonances} Transverse and longitudinal helicity amplitudes
         $A_{1/2}$, $A_{3/2}$ and $S_{1/2}$ for electromagnetic excitation
         of nucleon resonances off the proton at $Q^2 = 1.0$~GeV$^2$ in units
         of $10^{-3}\,GeV^{-1/2}$. The default values of MAID2000 are compared
         to our MAID2003 {\em local fit}. The $R_{EM} = E2/M1$
         and $R_{SM} = C2/M1$ ratios of the $\Delta(1232)$ are given in percentage.
         The errors given for the amplitudes are first the statistical errors
         of the fit and second the estimated model uncertainty. The errors of
         the ratios include both and are mainly model uncertainties. }
\begin{ruledtabular}
\begin{tabular}{crcl}
$N^*$ & & MAID2000 & MAID2003 \\
      & & default values & {\em local fit} \\
\hline $P_{33}(1232)$
   & $A_{1/2}$ &  -75 &   -70 $\pm$  1 $\pm$  2\\
   & $A_{3/2}$ & -142 &  -161 $\pm$  3 $\pm$  5\\
   & $S_{1/2}$ &   15 &    17 $\pm$  1 $\pm$  2\\
\hline
   & $R_{EM} $ & -2.2 &  -6.4 $\pm$  2.6       \\
   & $R_{SM} $ & -6.5 &  -7.0 $\pm$  1.7       \\
\hline $P_{11}(1440)$
   & $A_{1/2}$ &  -61 &    18 $\pm$  5 $\pm$ 20\\
   & $S_{1/2}$ &   20 &    19 $\pm$  3 $\pm$ 10\\
\hline $D_{13}(1520)$
   & $A_{1/2}$ &  -69 &   -77 $\pm$  7 $\pm$ 20\\
   & $A_{3/2}$ &   38 &    40 $\pm$  7 $\pm$ 10\\
   & $S_{1/2}$ &    0 &   -17 $\pm$  8 $\pm$ 10\\
\hline $S_{11}(1535)$
   & $A_{1/2}$ &   67 &    74 $\pm$ 10 $\pm$ 25\\
   & $S_{1/2}$ &    0 &   -22 $\pm$  5 $\pm$ 10\\
\hline $F_{15}(1680)$
   & $A_{1/2}$ &  -42 &   -36 $\pm$  5 $\pm$ 10\\
   & $A_{3/2}$ &   51 &    31 $\pm$ 10 $\pm$ 10\\
   & $S_{1/2}$ &    0 &   -22 $\pm$  5 $\pm$ 10\\
\end{tabular}
\end{ruledtabular}
\end{table}
\endgroup

%%%%%%%%%%%%%%%%%%%%%%%
\subsection{New Solution from SAID Analysis}
\label{subsec:SAID}
%%%%%%%%%%%%%%%%%%%%%%%

The predictions of the SAID analysis with  the {\em NF18K solution} parameter
set~\cite{Arndt:1996ak} are shown in Figs.~\ref{fig-wscan} and \ref{fig-phiscan}.
This is an extrapolation to the new kinematics of our
experiment.  When our data are included in the world dataset, a new
SAID fit, {\em WI03K solution}~\cite{Arndt:2003zk} yields
a much better fit, as shown in Fig.~\ref{fig-ctcmscan} and
Fig.~\ref{fig-sigma}.  We find a noticeable improvement
in the region of the $P_{33}(1232)$ resonance, resulting from improved
constraints on the $Q^2$ dependence of this resonance.  Although
$\sigma_{\mbox{\tiny TL}}$ is under-fitted at the $P_{33}$, in general SAID
{\em WI03K solution} gives an excellent description of $\sigma_{\mbox{\tiny TL}}$
and $\sigma_{\mbox{\tiny TT}}$ up to $W=1.7$~GeV. 

%%%%%%%%%%%%%%%%%%%%%%%%%%%%%%%%%%
\subsection{$\theta^*$ dependence}
\label{subsec:theta}
%%%%%%%%%%%%%%%%%%%%%%%%%%%%%%%%%%

\begin{figure}
\includegraphics[width=8.6cm]{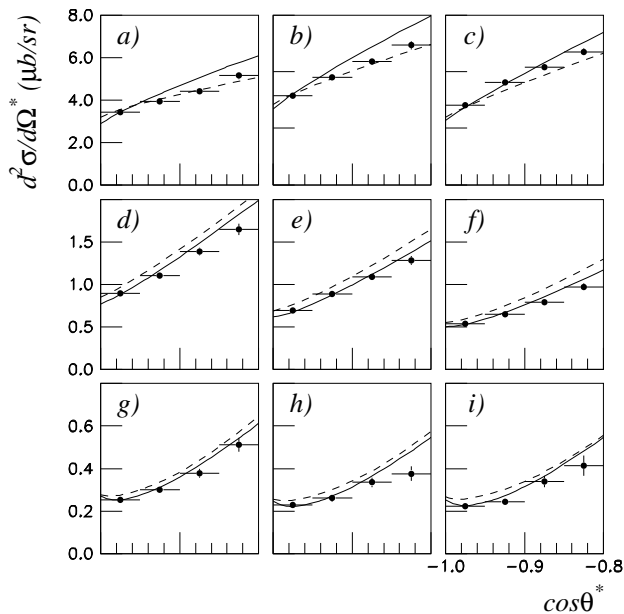}
\caption{\label{fig-ctcmscan}The cross section of reaction~(\ref{react1})
         as a function of $cos \theta^*$ obtained at $Q^2 = 1.0$~GeV$^2$,
         $\phi = 75^\circ$ for different values of $W$:
         1190~MeV --- a), 1210~MeV --- b), 1230~MeV --- c),
         1310~MeV --- d), 1330~MeV --- e), 1350~MeV --- f),
         1410~MeV --- g), 1430~MeV --- h) and 1450~MeV --- i).
         In all plots, $-1.0 \le \cos\theta^* \le -0.8$.
         The solid curve depicts the MAID2003 {\em local fit} and the
         dashed line to the SAID {\em WI03K solution}~\cite{Arndt:2003zk}.}
\end{figure}

The kinematic restrictions of the present experiment allow us to reliably
reconstruct the event distributions as a function of the pion angle
$\theta^\ast$ in the interval $-1 \leq cos \theta^\ast \leq -0.8$. The
corresponding cross section is shown in Fig.~\ref{fig-ctcmscan}
after optimization of the MAID and SAID parameters.

Overall, the relative shape in $\theta^*$ of the MAID2003 {\em local fit}
compared to the experimental data is good for all bins in $W$.
%One should notice that the normalization discrepancy observed in
%Fig.~\ref{fig-ctcmscan}i is due to the $W$ dependence of the cross
%section and not to the $\theta^*$ dependence.

%%%%% SECTION IV %%%%%%%%%%%%%%%%%%%%%%%%%%%%%%%%%%%%%%%%%%%%%%%%%%%%%%%%%%%%%%%%
\section{Study of the $Q^2$ dependence}
\label{sec:Q2dep}
%%%%%%%%%%%%%%%%%%%%%%%%%%%%%%%%%%%%%%%%%%%%%%%%%%%%%%%%%%%%%%%%%%%%%%%%%%%%%%%%%

We considered the correlation between $\phi$ and $Q^2$ due to
the acceptance as a possible source of systematic error. To minimize
this effect, we need a more realistic $Q^2$ dependence in the model.
To this end we first extract the experimental $Q^2$
dependence for each bin in $W$ and, second, iterate the analysis
using a model cross section $d\sigma_{\rm model}$ that includes
this dependence.

The cross section was evaluated by splitting our $Q^2$ range
[0.85,~1.15]~GeV$^2$ into six intervals, integrating over
cos$\theta^*$ in the range [--1.0,--0.8], and fitting the $\phi$
dependence of the cross section in a similar way to that described
in Section~\ref{sec:wphidep}.

\begin{figure}
\includegraphics[width=8.6cm]{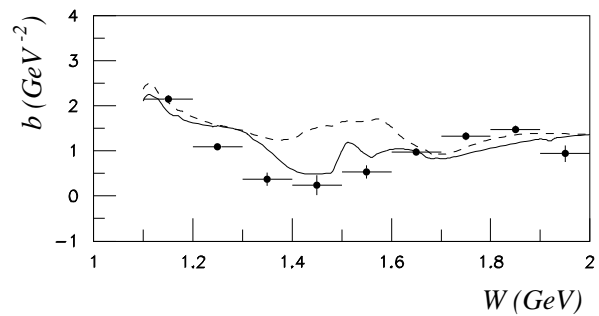}
\caption{\label{fig-q2dep}$Q^2$ evolution of
         $\sigma_{\mbox{\tiny T}}+\epsilon\ \sigma_{\mbox{\tiny L}}$
         evaluated at a central $Q^2$ = 1.0~GeV$^2$ from the data points.
         The dashed and solid curve correspond to the MAID2000 and MAID2003 {\em local fit}
         calculations respectively. See text for definition of $b$.}
\end{figure}

The $Q^2$ dependence of the cross section can be studied by
fitting the following form to the partial
$\sigma_{\mbox{\tiny T}}+\epsilon\ \sigma_{\mbox{\tiny L}}$ cross
section~:
\begin{eqnarray}
d\sigma(W,Q^2)=d\sigma(W,Q^2=1\,{\rm GeV}^2)\cdot e^{-b_{\rm exp}\cdot(1\,{\rm GeV}^2-Q^2)}.
\label{bfit}
\end{eqnarray}
The resulting fit values for $b_{\rm exp}$ are displayed in
Fig.~\ref{fig-q2dep}. We performed a similar exercise on the cross
section evaluated within the MAID2000 calculation and MAID2003 {\em local fit}.
The resulting parameter $b_{\rm maid}(W)$ is displayed in Fig.~\ref{fig-q2dep}
respectively with a full and a dashed curve.
While the overall ranges of variation of $b_{\rm exp}$ and $b_{\rm maid}$
are consistent, we observe a substantial discrepancy
between the model and the data in the range of $W$ from $\sim1.25$ to
$\sim1.65$~GeV.

The data obtained in the present analysis allow us to determine the
$Q^2$ dependence of the partial cross section
$\sigma_{\mbox{\tiny T}}+\epsilon\ \sigma_{\mbox{\tiny L}}$ over a wide
range of $W$.
%which at this time was not possible from other experiments.
In principle, it is also possible to study the $Q^2$ dependence of the
partial cross sections $\sigma_{\mbox{\tiny TT}}$ and $\sigma_{\mbox{\tiny TL}}$.
However, such an analysis would require much more statistics for a meaningful
interpretation.

The experimental $Q^2$ dependence of Eq.~\ref{bfit} is used in our
final data analysis with the MAID2003 {\em local fit} parameters.
In addition, we compute a systematic error associated with the $Q^2$-dependent
interpolation in the data analysis.
This systematic error is evaluated from one half of the difference
between the final analysis and the results obtained from the analysis
using the MAID2003 {\em local fit} without additional $Q^2$-dependence.

%%%%% SECTION V %%%%%%%%%%%%%%%%%%%%%%%%%%%%%%%%%%%%%%%%%%%%%%%%%%%%%%%%%%%%%%%%%
\section{DISCUSSION AND CONCLUSIONS}
\label{sec:results}
%%%%%%%%%%%%%%%%%%%%%%%%%%%%%%%%%%%%%%%%%%%%%%%%%%%%%%%%%%%%%%%%%%%%%%%%%%%%%%%%%

\begin{figure*}
\includegraphics[width=17.8cm,height=20cm]{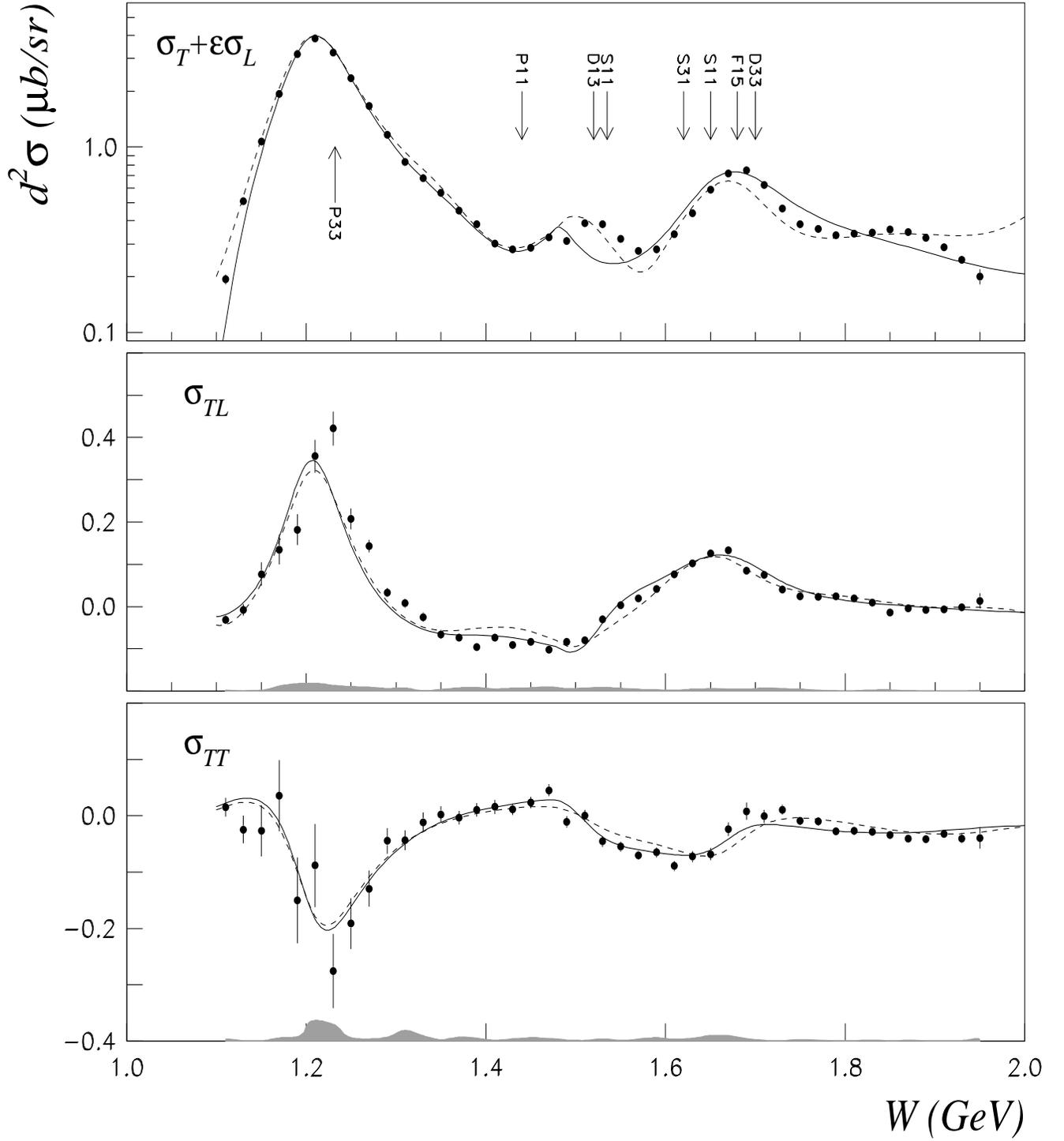}
\caption{\label{fig-sigma}Virtual photo-production cross sections
for $\gamma^* p \to p \pi^0$ with statistical error bars as
a function of $W$ at $Q^2$ = 1.0~GeV$^2$, $\cos{\theta^*} = -0.975$~:
$\sigma_{\mbox{\tiny T}}+\epsilon\ \sigma_{\mbox{\tiny L}}$,
$\sigma_{\mbox{\tiny TL}}$ and $\sigma_{\mbox{\tiny TT}}$.
The one-sigma value of the total systematic errors is given for the
$\sigma_{\mbox{\tiny TL}}$ and $\sigma_{\mbox{\tiny TT}}$ cross sections
by the size of the shaded area at the bottom of each plot.
The solid curves correspond to the MAID2003 {\em local fit}, and the dashed
curves to SAID {\em WI03K solution}~\cite{Arndt:2003zk}. In the top part of
the figure, we indicated the positions of the 8 most prominent resonances
whose helicity amplitudes are adjusted in MAID2003 (see Section~\ref{subsec:MAID}).}
\end{figure*}

\begin{figure}
\includegraphics[width=8.6cm]{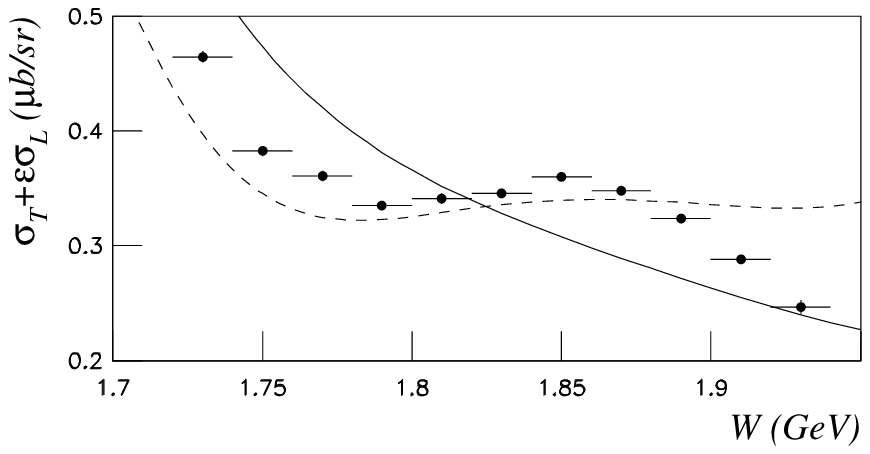}
\caption{\label{fig-zoom} The
         $\sigma_{\mbox{\tiny T}}+\epsilon\ \sigma_{\mbox{\tiny L}}$
         cross section in a limited high $W$ region at $Q^2 = 1.0$~GeV$^2$,
         $\cos{\theta^*} = --0.975$. The full line corresponds to the
         MAID2003 {\em local fit} and the dashed line to the SAID {\em WI03K solution}.}
\end{figure}

The final cross-section data are listed in
Tables~\ref{T+L_crossection},~\ref{TL_crossection}
and~\ref{TT_crossection}~\cite{tablespi0}. The value of
$\epsilon$ indicated in these tables corresponds to
a fixed value of $k=4032$~MeV within each considered
interval in $W$.

Figure~\ref{fig-sigma} presents the final cross sections
$\sigma_{\mbox{\tiny T}}+\epsilon\ \sigma_{\mbox{\tiny L}}$,
$\sigma_{\mbox{\tiny TL}}$ and $\sigma_{\mbox{\tiny TT}}$ as a function of $W$,
evaluated at $\cos{\theta^*}\ =\ -0.975$.

The systematic errors obtained from the iterative procedure
described in Section~\ref{sec:Q2dep} are added quadratically to the
errors listed in Table~\ref{systematic}. The total systematic error is
shown in Tables~\ref{T+L_crossection},\ref{TL_crossection},
\ref{TT_crossection} and in Fig.~\ref{fig-sigma}. It is of the
same order as the statistical error.

The parameter $b_{\rm exp}$ introduced in Section~\ref{sec:Q2dep}
can be phenomenologically related with the scale parameter $\Lambda$,
which determines the $Q^2$ dependence of hadronic form factors and
resonance multipoles in the dipole approximation:
\begin{equation}
G_{\rm D}(Q^2) = {1 \over (1 + Q^2/ \Lambda^2)^2}~,
\label{dipole}
\end{equation}
Assuming that $d^2 \sigma(Q^2) \sim( G_{\rm D}(Q^2))^2$ one finds
that $b_{\rm exp} \approx$ $4/(Q^2+\Lambda^2)$. Therefore
$b_{\rm exp}\rightarrow 0$ when the target is structureless
($\Lambda^2\rightarrow \infty$), $b_{\rm exp}\rightarrow 4/Q^2 = 4$~GeV$^{-2}$
at $Q^2 = 1$~GeV$^2$ in the case $\Lambda^2\ll Q^2$.
A standard fit to nucleon elastic form factors data yields $\Lambda^2=0.71$~GeV$^2$,
which at $Q^2=1$~GeV$^2$ corresponds to $b_{\rm exp} =2.3$~GeV$^{-2}$.

The range of observed variations of the parameter $b_{exp}$ displayed
in Fig.~\ref{fig-q2dep} lies essentially within the limits
$0<b_{\rm exp}<4$~GeV$^2$. Thus, it is consistent with the $Q^2$
dependence of the dipole form factor discussed above.

While Eq.~(\ref{dipole}) provides a reasonable approximation for
nucleon form factors in the range of $Q^2$ considered in this
study, it is known to deviate~\cite{Sterman:1997sx} from the results
obtained for the $\gamma {\rm N} \Delta$ transition
form factor $G_M^*(Q^2)$ which describes the dominant magnetic
dipole excitation of the $\Delta(1232)$.
In particular, $G_M^*(Q^2)$ falls off faster with $Q^2$ than the
dipole form factor indicating a magnetic radius of the
resonance state larger than that of the nucleon~\cite{Frolov:1998pw}.
Our results on $b_{\rm exp}$ yield new information on this topic
especially for $W>$ 1.7~GeV even if the contributions
from resonant and non-resonant amplitudes are not separated.

A nontrivial $W$ dependence of the parameter $b_{\rm exp}$ results from an
interplay between contributions of the resonant and non-resonant amplitudes.
The latter are known to be small around the $\Delta(1232)$ and increase
monotonically with $W$. We find from the data that $b_{\rm exp} \approx 0$
in the range of $1.3 < W < 1.4$~GeV, which indicates a cancellation
of the different $Q^2$ dependences of the resonant (especially $P_{11}$--Roper)
and non-resonant amplitudes.
The dominance of the resonant amplitude $M_{1+}$ in the range $W<$ 1.3~GeV
results in $b_{\rm exp}\approx1-2$~GeV$^{-2}$, and
$b_{\rm exp}\approx1-1.5$~GeV$^{-2}$ in the range $W>$ 1.6~GeV where
non-resonant terms start to dominate.

In summary, we have measured in the resonance region the three partial cross
sections $\sigma_{\mbox{\tiny T}}+\epsilon\ \sigma_{\mbox{\tiny L}}$,
$\sigma_{\mbox{\tiny TT}}$, and $\sigma_{\mbox{\tiny TL}}$ for the reaction
$\gamma^* p \to p \pi^0$ at $Q^2$=1~GeV$^2$ and backward angles. We have
obtained the $Q^2$ dependence for the cross section integrated over 
angle~$\phi$~:
$\int{d\sigma~ d\phi}=\sigma_{\mbox{\tiny T}}+\epsilon\ \sigma_{\mbox{\tiny L}}$.
These data will be used to constrain models. A first step was done
for the Unitary Isobar Model MAID2000~\cite{Drechsel:1998hk,Kamalov:2001yi}.
From this analysis, we find new constraints on the $R_{EM}$ and $R_{SM}$ ratios
of the $\Delta(1232)$ resonance (Table~\ref{resonances}).
In spite of the agreement between our data and the calculation, the $Q^2$ dependence
of the total cross section is not reproduced by MAID2000. On the other hand we observe
a substantial improvement in the $Q^2$ dependence shown in Fig.~\ref{fig-q2dep} when
our MAID2003 {\em local fit} is used.

The MAID and SAID analyses employ fundamentally different
techniques for describing the scattering amplitude. Our data
result in significant readjustments in the parameters of both
models, and throughout the entire $W$ range. Although the results
of the model calculations agreed initially very poorly with our data,
a joint analysis with the world dataset resulted in a much improved
description of the data. This is explained by evident limitations in
the kinematics of the pre-existing dataset for even the relatively simple
$\gamma^* p \rightarrow p \pi^0$ reaction, particularly at high $W$.
For both MAID and SAID, our data show strong sensitivity to the $P_{11}(1440)$
Roper resonance, as evidenced by the large changes in the new fits to this region
of the spectrum.

Finally, although our results are not sufficient to allow a full
partial-wave analysis in the high $W$ region (between 1.7 and
2.0~GeV), the position of the enhancement of
$\sigma_{\mbox{\tiny T}}+\epsilon\ \sigma_{\mbox{\tiny L}}$
(see Fig.~\ref{fig-zoom}) is fully consistent with the recent
analysis of Chen {\it et al.}~\cite{Chen:2002mn}. The dynamical
model used in~\cite{Chen:2002mn} implies that the third $S_{11}$
resonance should have a mass $1846\pm47$~MeV.
Evidences of missing resonances in this region have also
been shown in pion electroproduction at CLAS~\cite{Ripani:2003yw},
in kaon photoproduction at SAPHIR~\cite{Tran:1998qw} and
CLAS~\cite{McNabb:2003nf}, and in $\pi N~\to~\pi\eta$~\cite{Capstick:1999dg}.
All these recent publications demonstrate the interest of both
theoreticians and experimentalists in a detailed understanding
of the nucleon resonance region, and point out the need for
accurate data in meson electro- and photo-production.

The underlying physics of the nucleon resonances and the transition
to deep inelastic scattering is still under investigation.
Therefore, new data on exclusive processes as a function of both
$W$ and $Q^2$ are of great value.

%%%%%%%%%%%%%%%%%%%%%%%%%%%%%%%%%%%%%%%%%%%%%%%%%%%%%%%%%%%%%%%%%%%%%%%%%%%%%%%%%
\begin{acknowledgments}
We wish to acknowledge accelerator staff who delivered
the beam, as well as the Hall A technical staff.
We are greatful to the SAID group for useful discussions and for
providing the updated fits.
This work was supported by DOE contract DE-AC05-84ER40150 under
which the Southeastern Universities Research Association (SURA)
operates the Thomas Jefferson National Accelerator Facility. We
acknowledge additional grants from the US DOE and NSF, the French
Centre National de la Recherche Scientifique and Commissariat \`a
l'Energie Atomique, the Conseil R\'egional d'Auvergne, the
FWO-Flanders (Belgium) and the BOF-Gent University. 
\end{acknowledgments}
%%%%%%%%%%%%%%%%%%%%%%%%%%%%%%%%%%%%%%%%%%%%%%%%%%%%%%%%%%%%%%%%%%%%%%%%%%%%%%%%%

%%%%%%%%%%%%%%%%%%%%%%%%%%%%%%%%%%%%%%%%%%%%%%%%%%%%%%%%%%%%%%%%%%%%%%%%%%%%%%%%%
\bibliography{common}

%%%%%%%%%%%%%%%%%%%%%%%%%%%%%%%%%%%%%%%%%%%%%%%%%%%%%%%%%%%%%%%%%%%%%%%%%%%%%%%%%
\begin{turnpage}
\begingroup
\squeezetable
\begin{table*}
\caption{\label{T+L_crossection}$\sigma_{\mbox{\tiny T}}+\epsilon\ \sigma_{\mbox{\tiny L}}$
         cross section at $Q^2$ = 1.0~GeV$^2$ in $\mu b.sr^{-1}$. The values
         are followed by the statistical and the total systematic errors.}
\begin{ruledtabular}
\begin{tabular}{|c|c|ccc|ccc|ccc|ccc|} \hline
$W$ (MeV)&$\epsilon$ &\multicolumn{3}{c|}{$\cos\theta^*=-0.975$}
&\multicolumn{3}{c|}{$\cos\theta^*=-0.925$}
&\multicolumn{3}{c|}{$\cos\theta^*=-0.875$}
&\multicolumn{3}{c|}{$\cos\theta^*=-0.825$} \\ \hline
$1110.0$&$0.945$&$ 0.194$&$\pm  0.012$&$\pm  0.006$&$ 0.241$&$\pm  0.018$&$\pm  0.008$&$ 0.300$&$\pm  0.028$&$\pm  0.010$&$ 0.313$&$\pm  0.045$&$\pm  0.010$ \\
$1130.0$&$0.944$&$ 0.511$&$\pm  0.017$&$\pm  0.017$&$ 0.597$&$\pm  0.037$&$\pm  0.020$&$ 0.668$&$\pm  0.067$&$\pm  0.022$&$ 0.640$&$\pm  0.084$&$\pm  0.021$ \\
$1150.0$&$0.942$&$ 1.068$&$\pm  0.034$&$\pm  0.035$&$ 0.965$&$\pm  0.086$&$\pm  0.032$&$ 0.933$&$\pm  0.105$&$\pm  0.031$&$ 0.996$&$\pm  0.116$&$\pm  0.034$ \\
$1170.0$&$0.940$&$ 1.937$&$\pm  0.046$&$\pm  0.065$&$ 2.365$&$\pm  0.094$&$\pm  0.080$&$ 1.625$&$\pm  0.139$&$\pm  0.054$&$ 1.785$&$\pm  0.176$&$\pm  0.061$ \\
$1190.0$&$0.938$&$ 3.176$&$\pm  0.049$&$\pm  0.108$&$ 3.591$&$\pm  0.102$&$\pm  0.124$&$ 3.518$&$\pm  0.167$&$\pm  0.118$&$ 3.074$&$\pm  0.186$&$\pm  0.101$ \\
$1210.0$&$0.936$&$ 3.853$&$\pm  0.049$&$\pm  0.134$&$ 3.887$&$\pm  0.125$&$\pm  0.135$&$ 3.824$&$\pm  0.153$&$\pm  0.134$&$ 3.573$&$\pm  0.172$&$\pm  0.133$ \\
$1230.0$&$0.934$&$ 3.221$&$\pm  0.048$&$\pm  0.109$&$ 3.171$&$\pm  0.130$&$\pm  0.105$&$ 3.385$&$\pm  0.153$&$\pm  0.115$&$ 3.687$&$\pm  0.175$&$\pm  0.125$ \\
$1250.0$&$0.932$&$ 2.348$&$\pm  0.033$&$\pm  0.080$&$ 2.645$&$\pm  0.064$&$\pm  0.088$&$ 2.287$&$\pm  0.155$&$\pm  0.076$&$ 2.451$&$\pm  0.177$&$\pm  0.082$ \\
$1270.0$&$0.930$&$ 1.665$&$\pm  0.021$&$\pm  0.056$&$ 1.832$&$\pm  0.039$&$\pm  0.061$&$ 1.828$&$\pm  0.110$&$\pm  0.060$&$ 1.602$&$\pm  0.181$&$\pm  0.054$ \\
$1290.0$&$0.927$&$ 1.162$&$\pm  0.015$&$\pm  0.039$&$ 1.302$&$\pm  0.038$&$\pm  0.045$&$ 1.307$&$\pm  0.088$&$\pm  0.044$&$ 1.155$&$\pm  0.132$&$\pm  0.039$ \\
$1310.0$&$0.925$&$ 0.832$&$\pm  0.012$&$\pm  0.029$&$ 0.893$&$\pm  0.042$&$\pm  0.037$&$ 0.912$&$\pm  0.078$&$\pm  0.041$&$ 0.973$&$\pm  0.123$&$\pm  0.042$ \\
$1330.0$&$0.922$&$ 0.680$&$\pm  0.013$&$\pm  0.023$&$ 0.686$&$\pm  0.040$&$\pm  0.023$&$ 0.770$&$\pm  0.073$&$\pm  0.031$&$ 0.681$&$\pm  0.118$&$\pm  0.023$ \\
$1350.0$&$0.920$&$ 0.566$&$\pm  0.012$&$\pm  0.019$&$ 0.641$&$\pm  0.034$&$\pm  0.021$&$ 0.643$&$\pm  0.078$&$\pm  0.022$&$ 0.611$&$\pm  0.138$&$\pm  0.027$ \\
$1370.0$&$0.917$&$ 0.455$&$\pm  0.010$&$\pm  0.016$&$ 0.501$&$\pm  0.034$&$\pm  0.017$&$ 0.414$&$\pm  0.080$&$\pm  0.024$&$ 0.147$&$\pm  0.131$&$\pm  0.007$ \\
$1390.0$&$0.914$&$ 0.385$&$\pm  0.010$&$\pm  0.013$&$ 0.375$&$\pm  0.036$&$\pm  0.021$&$ 0.387$&$\pm  0.086$&$\pm  0.032$&$ 0.128$&$\pm  0.161$&$\pm  0.008$ \\
$1410.0$&$0.910$&$ 0.301$&$\pm  0.010$&$\pm  0.010$&$ 0.349$&$\pm  0.036$&$\pm  0.021$&$ 0.309$&$\pm  0.079$&$\pm  0.025$&$ 0.023$&$\pm  0.122$&$\pm  0.013$ \\
$1430.0$&$0.907$&$ 0.281$&$\pm  0.008$&$\pm  0.011$&$ 0.344$&$\pm  0.031$&$\pm  0.016$&$ 0.441$&$\pm  0.102$&$\pm  0.018$&$ 0.160$&$\pm  0.151$&$\pm  0.009$ \\
$1450.0$&$0.903$&$ 0.286$&$\pm  0.007$&$\pm  0.011$&$ 0.371$&$\pm  0.029$&$\pm  0.019$&$ 0.271$&$\pm  0.087$&$\pm  0.014$&$-0.002$&$\pm  0.171$&$\pm  0.009$ \\
$1470.0$&$0.900$&$ 0.327$&$\pm  0.008$&$\pm  0.012$&$ 0.329$&$\pm  0.036$&$\pm  0.019$&$ 0.255$&$\pm  0.132$&$\pm  0.024$&&& \\
$1490.0$&$0.896$&$ 0.311$&$\pm  0.008$&$\pm  0.011$&$ 0.460$&$\pm  0.048$&$\pm  0.029$&$ 0.364$&$\pm  0.123$&$\pm  0.030$&&& \\
$1510.0$&$0.892$&$ 0.388$&$\pm  0.008$&$\pm  0.013$&$ 0.430$&$\pm  0.036$&$\pm  0.015$&$ 0.314$&$\pm  0.114$&$\pm  0.011$&&& \\
$1530.0$&$0.887$&$ 0.383$&$\pm  0.007$&$\pm  0.013$&$ 0.488$&$\pm  0.032$&$\pm  0.016$&$ 0.409$&$\pm  0.135$&$\pm  0.014$&&& \\
$1550.0$&$0.883$&$ 0.319$&$\pm  0.006$&$\pm  0.011$&$ 0.396$&$\pm  0.051$&$\pm  0.021$&$-0.111$&$\pm  0.168$&$\pm  0.013$&&& \\
$1570.0$&$0.878$&$ 0.276$&$\pm  0.005$&$\pm  0.009$&$ 0.461$&$\pm  0.059$&$\pm  0.032$&$-0.176$&$\pm  0.214$&$\pm  0.009$&&& \\
$1590.0$&$0.873$&$ 0.281$&$\pm  0.005$&$\pm  0.011$&$ 0.376$&$\pm  0.055$&$\pm  0.013$&$-0.107$&$\pm  0.142$&$\pm  0.019$&&& \\
$1610.0$&$0.868$&$ 0.339$&$\pm  0.005$&$\pm  0.013$&$ 0.428$&$\pm  0.040$&$\pm  0.018$&$-0.112$&$\pm  0.157$&$\pm  0.004$&&& \\
$1630.0$&$0.863$&$ 0.439$&$\pm  0.006$&$\pm  0.018$&$ 0.450$&$\pm  0.064$&$\pm  0.020$&&&&&& \\
$1650.0$&$0.857$&$ 0.590$&$\pm  0.006$&$\pm  0.028$&$ 0.602$&$\pm  0.085$&$\pm  0.040$&&&&&& \\
$1670.0$&$0.852$&$ 0.719$&$\pm  0.008$&$\pm  0.041$&$ 0.606$&$\pm  0.085$&$\pm  0.046$&&&&&& \\
$1690.0$&$0.845$&$ 0.749$&$\pm  0.010$&$\pm  0.048$&$ 0.535$&$\pm  0.106$&$\pm  0.018$&&&&&& \\
$1710.0$&$0.839$&$ 0.625$&$\pm  0.007$&$\pm  0.035$&$-0.069$&$\pm  0.187$&$\pm  0.003$&&&&&& \\
$1730.0$&$0.832$&$ 0.465$&$\pm  0.005$&$\pm  0.024$&$ 0.348$&$\pm  0.063$&$\pm  0.022$&&&&&& \\
$1750.0$&$0.826$&$ 0.383$&$\pm  0.004$&$\pm  0.016$&$ 0.281$&$\pm  0.042$&$\pm  0.015$&&&&&& \\
$1770.0$&$0.818$&$ 0.361$&$\pm  0.005$&$\pm  0.013$&$ 0.082$&$\pm  0.079$&$\pm  0.007$&&&&&& \\
$1790.0$&$0.811$&$ 0.335$&$\pm  0.004$&$\pm  0.011$&&&&&&&&& \\
$1810.0$&$0.803$&$ 0.341$&$\pm  0.005$&$\pm  0.011$&&&&&&&&& \\
$1830.0$&$0.795$&$ 0.346$&$\pm  0.004$&$\pm  0.011$&&&&&&&&& \\
$1850.0$&$0.786$&$ 0.360$&$\pm  0.004$&$\pm  0.012$&&&&&&&&& \\
$1870.0$&$0.777$&$ 0.348$&$\pm  0.004$&$\pm  0.011$&&&&&&&&& \\
$1890.0$&$0.768$&$ 0.323$&$\pm  0.004$&$\pm  0.011$&&&&&&&&& \\
$1910.0$&$0.759$&$ 0.288$&$\pm  0.004$&$\pm  0.010$&&&&&&&&& \\
$1930.0$&$0.749$&$ 0.247$&$\pm  0.006$&$\pm  0.008$&&&&&&&&& \\
$1950.0$&$0.738$&$ 0.199$&$\pm  0.019$&$\pm  0.007$&&&&&&&&& \\
\hline \end{tabular}
\end{ruledtabular}
\end{table*}
\endgroup
\end{turnpage}

\begin{turnpage}
\begingroup
\squeezetable
\begin{table*}
\caption{\label{TL_crossection}$\sigma_{\mbox{\tiny TL}}$
         cross section at $Q^2$ = 1.0~GeV$^2$ in $\mu b.sr^{-1}$. The values
         are followed by the statistical and the total systematic errors.}
\begin{ruledtabular}
\begin{tabular}{|c|c|ccc|ccc|ccc|ccc|} \hline
$W$ (MeV)&$\epsilon$ &\multicolumn{3}{c|}{$\cos\theta^*=-0.975$}
&\multicolumn{3}{c|}{$\cos\theta^*=-0.925$}
&\multicolumn{3}{c|}{$\cos\theta^*=-0.875$}
&\multicolumn{3}{c|}{$\cos\theta^*=-0.825$} \\ \hline
$1110.0$&$0.945$&$-0.032$&$\pm  0.009$&$\pm  0.001$&$-0.087$&$\pm  0.015$&$\pm  0.003$&$-0.076$&$\pm  0.024$&$\pm  0.003$&$-0.079$&$\pm  0.038$&$\pm  0.003$ \\
$1130.0$&$0.944$&$-0.008$&$\pm  0.013$&$\pm  0.001$&$-0.039$&$\pm  0.032$&$\pm  0.002$&$-0.045$&$\pm  0.058$&$\pm  0.002$&$ 0.003$&$\pm  0.071$&$\pm  0.003$ \\
$1150.0$&$0.942$&$ 0.077$&$\pm  0.028$&$\pm  0.003$&$ 0.216$&$\pm  0.074$&$\pm  0.007$&$ 0.413$&$\pm  0.089$&$\pm  0.014$&$ 0.400$&$\pm  0.097$&$\pm  0.014$ \\
$1170.0$&$0.940$&$ 0.135$&$\pm  0.035$&$\pm  0.012$&$ 0.094$&$\pm  0.076$&$\pm  0.020$&$ 0.896$&$\pm  0.114$&$\pm  0.031$&$ 0.986$&$\pm  0.142$&$\pm  0.033$ \\
$1190.0$&$0.938$&$ 0.182$&$\pm  0.037$&$\pm  0.018$&$ 0.418$&$\pm  0.087$&$\pm  0.033$&$ 0.879$&$\pm  0.137$&$\pm  0.033$&$ 1.529$&$\pm  0.148$&$\pm  0.051$ \\
$1210.0$&$0.936$&$ 0.356$&$\pm  0.039$&$\pm  0.019$&$ 0.922$&$\pm  0.110$&$\pm  0.042$&$ 1.619$&$\pm  0.129$&$\pm  0.055$&$ 2.072$&$\pm  0.139$&$\pm  0.069$ \\
$1230.0$&$0.934$&$ 0.421$&$\pm  0.040$&$\pm  0.014$&$ 1.141$&$\pm  0.111$&$\pm  0.040$&$ 1.537$&$\pm  0.128$&$\pm  0.054$&$ 1.796$&$\pm  0.141$&$\pm  0.062$ \\
$1250.0$&$0.932$&$ 0.208$&$\pm  0.025$&$\pm  0.011$&$ 0.502$&$\pm  0.052$&$\pm  0.025$&$ 1.182$&$\pm  0.128$&$\pm  0.046$&$ 1.489$&$\pm  0.143$&$\pm  0.056$ \\
$1270.0$&$0.930$&$ 0.143$&$\pm  0.015$&$\pm  0.010$&$ 0.293$&$\pm  0.033$&$\pm  0.016$&$ 0.601$&$\pm  0.091$&$\pm  0.025$&$ 1.112$&$\pm  0.146$&$\pm  0.049$ \\
$1290.0$&$0.927$&$ 0.033$&$\pm  0.011$&$\pm  0.007$&$ 0.201$&$\pm  0.033$&$\pm  0.014$&$ 0.426$&$\pm  0.074$&$\pm  0.022$&$ 0.773$&$\pm  0.109$&$\pm  0.035$ \\
$1310.0$&$0.925$&$ 0.008$&$\pm  0.010$&$\pm  0.008$&$ 0.129$&$\pm  0.037$&$\pm  0.031$&$ 0.359$&$\pm  0.061$&$\pm  0.038$&$ 0.467$&$\pm  0.092$&$\pm  0.042$ \\
$1330.0$&$0.922$&$-0.025$&$\pm  0.011$&$\pm  0.001$&$ 0.067$&$\pm  0.035$&$\pm  0.009$&$ 0.146$&$\pm  0.059$&$\pm  0.031$&$ 0.392$&$\pm  0.085$&$\pm  0.027$ \\
$1350.0$&$0.920$&$-0.066$&$\pm  0.010$&$\pm  0.004$&$-0.056$&$\pm  0.029$&$\pm  0.009$&$ 0.044$&$\pm  0.064$&$\pm  0.022$&$ 0.164$&$\pm  0.104$&$\pm  0.015$ \\
$1370.0$&$0.917$&$-0.073$&$\pm  0.008$&$\pm  0.008$&$-0.067$&$\pm  0.027$&$\pm  0.011$&$ 0.094$&$\pm  0.066$&$\pm  0.027$&$ 0.387$&$\pm  0.104$&$\pm  0.019$ \\
$1390.0$&$0.914$&$-0.096$&$\pm  0.008$&$\pm  0.008$&$-0.038$&$\pm  0.028$&$\pm  0.018$&$ 0.001$&$\pm  0.062$&$\pm  0.025$&$ 0.205$&$\pm  0.115$&$\pm  0.009$ \\
$1410.0$&$0.910$&$-0.073$&$\pm  0.009$&$\pm  0.005$&$-0.088$&$\pm  0.029$&$\pm  0.021$&$-0.004$&$\pm  0.058$&$\pm  0.026$&$ 0.233$&$\pm  0.089$&$\pm  0.019$ \\
$1430.0$&$0.907$&$-0.091$&$\pm  0.007$&$\pm  0.007$&$-0.110$&$\pm  0.026$&$\pm  0.017$&$-0.140$&$\pm  0.076$&$\pm  0.020$&$ 0.125$&$\pm  0.113$&$\pm  0.008$ \\
$1450.0$&$0.903$&$-0.084$&$\pm  0.006$&$\pm  0.008$&$-0.155$&$\pm  0.024$&$\pm  0.020$&$-0.005$&$\pm  0.070$&$\pm  0.021$&$ 0.238$&$\pm  0.130$&$\pm  0.013$ \\
$1470.0$&$0.900$&$-0.102$&$\pm  0.007$&$\pm  0.011$&$-0.101$&$\pm  0.026$&$\pm  0.016$&$-0.046$&$\pm  0.071$&$\pm  0.017$&&& \\
$1490.0$&$0.896$&$-0.084$&$\pm  0.007$&$\pm  0.006$&$-0.138$&$\pm  0.034$&$\pm  0.030$&$-0.032$&$\pm  0.076$&$\pm  0.031$&&& \\
$1510.0$&$0.892$&$-0.079$&$\pm  0.007$&$\pm  0.008$&$-0.056$&$\pm  0.029$&$\pm  0.018$&$ 0.083$&$\pm  0.078$&$\pm  0.019$&&& \\
$1530.0$&$0.887$&$-0.030$&$\pm  0.005$&$\pm  0.009$&$-0.017$&$\pm  0.026$&$\pm  0.020$&$ 0.092$&$\pm  0.092$&$\pm  0.022$&&& \\
$1550.0$&$0.883$&$ 0.003$&$\pm  0.005$&$\pm  0.007$&$ 0.030$&$\pm  0.030$&$\pm  0.029$&$ 0.277$&$\pm  0.092$&$\pm  0.023$&&& \\
$1570.0$&$0.878$&$ 0.019$&$\pm  0.004$&$\pm  0.005$&$ 0.047$&$\pm  0.036$&$\pm  0.036$&$ 0.266$&$\pm  0.097$&$\pm  0.029$&&& \\
$1590.0$&$0.873$&$ 0.041$&$\pm  0.005$&$\pm  0.003$&$ 0.098$&$\pm  0.037$&$\pm  0.022$&$ 0.180$&$\pm  0.104$&$\pm  0.021$&&& \\
$1610.0$&$0.868$&$ 0.076$&$\pm  0.005$&$\pm  0.003$&$ 0.127$&$\pm  0.029$&$\pm  0.006$&$ 0.390$&$\pm  0.105$&$\pm  0.015$&&& \\
$1630.0$&$0.863$&$ 0.102$&$\pm  0.005$&$\pm  0.005$&$ 0.157$&$\pm  0.036$&$\pm  0.007$&&&&&& \\
$1650.0$&$0.857$&$ 0.125$&$\pm  0.005$&$\pm  0.006$&$ 0.217$&$\pm  0.046$&$\pm  0.024$&&&&&& \\
$1670.0$&$0.852$&$ 0.133$&$\pm  0.007$&$\pm  0.005$&$ 0.339$&$\pm  0.050$&$\pm  0.026$&&&&&& \\
$1690.0$&$0.845$&$ 0.085$&$\pm  0.009$&$\pm  0.005$&$ 0.198$&$\pm  0.051$&$\pm  0.008$&&&&&& \\
$1710.0$&$0.839$&$ 0.075$&$\pm  0.006$&$\pm  0.008$&$ 0.092$&$\pm  0.054$&$\pm  0.005$&&&&&& \\
$1730.0$&$0.832$&$ 0.041$&$\pm  0.004$&$\pm  0.006$&$ 0.117$&$\pm  0.037$&$\pm  0.017$&&&&&& \\
$1750.0$&$0.826$&$ 0.025$&$\pm  0.004$&$\pm  0.005$&$ 0.031$&$\pm  0.025$&$\pm  0.011$&&&&&& \\
$1770.0$&$0.818$&$ 0.023$&$\pm  0.004$&$\pm  0.003$&$ 0.064$&$\pm  0.047$&$\pm  0.015$&&&&&& \\
$1790.0$&$0.811$&$ 0.024$&$\pm  0.004$&$\pm  0.001$&&&&&&&&& \\
$1810.0$&$0.803$&$ 0.020$&$\pm  0.004$&$\pm  0.001$&&&&&&&&& \\
$1830.0$&$0.795$&$ 0.010$&$\pm  0.004$&$\pm  0.003$&&&&&&&&& \\
$1850.0$&$0.786$&$-0.014$&$\pm  0.003$&$\pm  0.003$&&&&&&&&& \\
$1870.0$&$0.777$&$-0.004$&$\pm  0.004$&$\pm  0.001$&&&&&&&&& \\
$1890.0$&$0.768$&$-0.008$&$\pm  0.004$&$\pm  0.000$&&&&&&&&& \\
$1910.0$&$0.759$&$-0.006$&$\pm  0.004$&$\pm  0.000$&&&&&&&&& \\
$1930.0$&$0.749$&$-0.002$&$\pm  0.006$&$\pm  0.000$&&&&&&&&& \\
$1950.0$&$0.738$&$ 0.015$&$\pm  0.018$&$\pm  0.001$&&&&&&&&& \\
\hline \end{tabular}
\end{ruledtabular}
\end{table*}
\endgroup
\end{turnpage}

\begin{turnpage}
\begingroup
\squeezetable
\begin{table*}
\caption{\label{TT_crossection}$\sigma_{\mbox{\tiny TT}}$
         cross section at $Q^2$ = 1.0~GeV$^2$ in $\mu b.sr^{-1}$. The values
         are followed by the statistical and the total systematic errors.}
\begin{ruledtabular}
\begin{tabular}{|c|c|ccc|ccc|ccc|ccc|} \hline
$W$ (MeV)&$\epsilon$ &\multicolumn{3}{c|}{$\cos\theta^*=-0.975$}
&\multicolumn{3}{c|}{$\cos\theta^*=-0.925$}
&\multicolumn{3}{c|}{$\cos\theta^*=-0.875$}
&\multicolumn{3}{c|}{$\cos\theta^*=-0.825$} \\ \hline
$1110.0$&$0.945$&$ 0.015$&$\pm  0.016$&$\pm  0.001$&$ 0.039$&$\pm  0.021$&$\pm  0.002$&$ 0.044$&$\pm  0.034$&$\pm  0.002$&$ 0.002$&$\pm  0.048$&$\pm  0.001$ \\
$1130.0$&$0.944$&$-0.024$&$\pm  0.025$&$\pm  0.001$&$-0.016$&$\pm  0.044$&$\pm  0.001$&$-0.071$&$\pm  0.068$&$\pm  0.002$&$-0.034$&$\pm  0.079$&$\pm  0.002$ \\
$1150.0$&$0.942$&$-0.026$&$\pm  0.045$&$\pm  0.001$&$-0.269$&$\pm  0.086$&$\pm  0.009$&$-0.445$&$\pm  0.097$&$\pm  0.015$&$-0.437$&$\pm  0.105$&$\pm  0.015$ \\
$1170.0$&$0.940$&$ 0.035$&$\pm  0.063$&$\pm  0.006$&$ 0.040$&$\pm  0.094$&$\pm  0.014$&$-0.845$&$\pm  0.124$&$\pm  0.028$&$-1.194$&$\pm  0.141$&$\pm  0.040$ \\
$1190.0$&$0.938$&$-0.150$&$\pm  0.076$&$\pm  0.008$&$-0.358$&$\pm  0.127$&$\pm  0.014$&$-0.917$&$\pm  0.164$&$\pm  0.031$&$-1.825$&$\pm  0.177$&$\pm  0.064$ \\
$1210.0$&$0.936$&$-0.088$&$\pm  0.074$&$\pm  0.037$&$-0.836$&$\pm  0.146$&$\pm  0.043$&$-1.644$&$\pm  0.170$&$\pm  0.063$&$-2.342$&$\pm  0.201$&$\pm  0.084$ \\
$1230.0$&$0.934$&$-0.275$&$\pm  0.066$&$\pm  0.030$&$-1.152$&$\pm  0.135$&$\pm  0.079$&$-1.611$&$\pm  0.151$&$\pm  0.083$&$-1.897$&$\pm  0.184$&$\pm  0.086$ \\
$1250.0$&$0.932$&$-0.191$&$\pm  0.045$&$\pm  0.007$&$-0.557$&$\pm  0.072$&$\pm  0.039$&$-1.117$&$\pm  0.133$&$\pm  0.061$&$-1.507$&$\pm  0.155$&$\pm  0.075$ \\
$1270.0$&$0.930$&$-0.129$&$\pm  0.032$&$\pm  0.004$&$-0.391$&$\pm  0.054$&$\pm  0.022$&$-0.668$&$\pm  0.102$&$\pm  0.040$&$-1.283$&$\pm  0.143$&$\pm  0.050$ \\
$1290.0$&$0.927$&$-0.045$&$\pm  0.023$&$\pm  0.007$&$-0.256$&$\pm  0.051$&$\pm  0.009$&$-0.643$&$\pm  0.093$&$\pm  0.025$&$-0.948$&$\pm  0.128$&$\pm  0.035$ \\
$1310.0$&$0.925$&$-0.043$&$\pm  0.018$&$\pm  0.019$&$-0.174$&$\pm  0.050$&$\pm  0.007$&$-0.388$&$\pm  0.094$&$\pm  0.013$&$-0.598$&$\pm  0.142$&$\pm  0.020$ \\
$1330.0$&$0.922$&$-0.011$&$\pm  0.017$&$\pm  0.010$&$-0.168$&$\pm  0.042$&$\pm  0.023$&$-0.267$&$\pm  0.075$&$\pm  0.012$&$-0.508$&$\pm  0.135$&$\pm  0.041$ \\
$1350.0$&$0.920$&$ 0.002$&$\pm  0.015$&$\pm  0.003$&$-0.034$&$\pm  0.032$&$\pm  0.015$&$-0.109$&$\pm  0.068$&$\pm  0.012$&$-0.321$&$\pm  0.128$&$\pm  0.054$ \\
$1370.0$&$0.917$&$-0.004$&$\pm  0.012$&$\pm  0.007$&$-0.003$&$\pm  0.033$&$\pm  0.001$&$-0.233$&$\pm  0.068$&$\pm  0.008$&$-0.516$&$\pm  0.116$&$\pm  0.027$ \\
$1390.0$&$0.914$&$ 0.011$&$\pm  0.012$&$\pm  0.005$&$-0.070$&$\pm  0.037$&$\pm  0.021$&$-0.151$&$\pm  0.084$&$\pm  0.031$&$-0.537$&$\pm  0.164$&$\pm  0.019$ \\
$1410.0$&$0.910$&$ 0.016$&$\pm  0.012$&$\pm  0.002$&$-0.004$&$\pm  0.034$&$\pm  0.019$&$-0.132$&$\pm  0.078$&$\pm  0.028$&$-0.516$&$\pm  0.123$&$\pm  0.025$ \\
$1430.0$&$0.907$&$ 0.012$&$\pm  0.010$&$\pm  0.004$&$ 0.040$&$\pm  0.029$&$\pm  0.009$&$ 0.003$&$\pm  0.095$&$\pm  0.016$&$-0.258$&$\pm  0.139$&$\pm  0.010$ \\
$1450.0$&$0.903$&$ 0.023$&$\pm  0.010$&$\pm  0.006$&$ 0.091$&$\pm  0.029$&$\pm  0.009$&$-0.114$&$\pm  0.080$&$\pm  0.006$&$-0.466$&$\pm  0.160$&$\pm  0.018$ \\
$1470.0$&$0.900$&$ 0.045$&$\pm  0.011$&$\pm  0.005$&$ 0.013$&$\pm  0.043$&$\pm  0.015$&$-0.163$&$\pm  0.156$&$\pm  0.022$&&& \\
$1490.0$&$0.896$&$-0.011$&$\pm  0.010$&$\pm  0.002$&$ 0.069$&$\pm  0.055$&$\pm  0.028$&$-0.182$&$\pm  0.143$&$\pm  0.036$&&& \\
$1510.0$&$0.892$&$ 0.000$&$\pm  0.011$&$\pm  0.002$&$-0.057$&$\pm  0.042$&$\pm  0.005$&$-0.298$&$\pm  0.132$&$\pm  0.010$&&& \\
$1530.0$&$0.887$&$-0.045$&$\pm  0.010$&$\pm  0.004$&$-0.071$&$\pm  0.040$&$\pm  0.004$&$-0.221$&$\pm  0.156$&$\pm  0.008$&&& \\
$1550.0$&$0.883$&$-0.055$&$\pm  0.009$&$\pm  0.002$&$-0.196$&$\pm  0.067$&$\pm  0.016$&$-0.914$&$\pm  0.216$&$\pm  0.031$&&& \\
$1570.0$&$0.878$&$-0.070$&$\pm  0.008$&$\pm  0.004$&$-0.105$&$\pm  0.075$&$\pm  0.037$&$-0.883$&$\pm  0.282$&$\pm  0.030$&&& \\
$1590.0$&$0.873$&$-0.065$&$\pm  0.008$&$\pm  0.006$&$-0.210$&$\pm  0.072$&$\pm  0.011$&$-0.817$&$\pm  0.164$&$\pm  0.049$&&& \\
$1610.0$&$0.868$&$-0.089$&$\pm  0.009$&$\pm  0.004$&$-0.228$&$\pm  0.054$&$\pm  0.011$&$-0.756$&$\pm  0.206$&$\pm  0.025$&&& \\
$1630.0$&$0.863$&$-0.072$&$\pm  0.010$&$\pm  0.005$&$-0.375$&$\pm  0.091$&$\pm  0.012$&&&&&& \\
$1650.0$&$0.857$&$-0.068$&$\pm  0.011$&$\pm  0.010$&$-0.225$&$\pm  0.114$&$\pm  0.017$&&&&&& \\
$1670.0$&$0.852$&$-0.024$&$\pm  0.012$&$\pm  0.010$&$-0.271$&$\pm  0.120$&$\pm  0.011$&&&&&& \\
$1690.0$&$0.845$&$ 0.008$&$\pm  0.015$&$\pm  0.005$&$-0.182$&$\pm  0.147$&$\pm  0.050$&&&&&& \\
$1710.0$&$0.839$&$-0.001$&$\pm  0.011$&$\pm  0.002$&$-0.829$&$\pm  0.260$&$\pm  0.050$&&&&&& \\
$1730.0$&$0.832$&$ 0.011$&$\pm  0.009$&$\pm  0.001$&$ 0.012$&$\pm  0.097$&$\pm  0.000$&&&&&& \\
$1750.0$&$0.826$&$-0.009$&$\pm  0.007$&$\pm  0.003$&$ 0.030$&$\pm  0.063$&$\pm  0.002$&&&&&& \\
$1770.0$&$0.818$&$-0.010$&$\pm  0.007$&$\pm  0.002$&$-0.318$&$\pm  0.124$&$\pm  0.024$&&&&&& \\
$1790.0$&$0.811$&$-0.027$&$\pm  0.006$&$\pm  0.001$&&&&&&&&& \\
$1810.0$&$0.803$&$-0.026$&$\pm  0.007$&$\pm  0.001$&&&&&&&&& \\
$1830.0$&$0.795$&$-0.028$&$\pm  0.006$&$\pm  0.002$&&&&&&&&& \\
$1850.0$&$0.786$&$-0.034$&$\pm  0.007$&$\pm  0.001$&&&&&&&&& \\
$1870.0$&$0.777$&$-0.041$&$\pm  0.007$&$\pm  0.002$&&&&&&&&& \\
$1890.0$&$0.768$&$-0.041$&$\pm  0.007$&$\pm  0.001$&&&&&&&&& \\
$1910.0$&$0.759$&$-0.033$&$\pm  0.007$&$\pm  0.001$&&&&&&&&& \\
$1930.0$&$0.749$&$-0.041$&$\pm  0.008$&$\pm  0.002$&&&&&&&&& \\
$1950.0$&$0.738$&$-0.041$&$\pm  0.019$&$\pm  0.002$&&&&&&&&& \\
\hline \end{tabular}
\end{ruledtabular}
\end{table*}
\endgroup
\end{turnpage}

\end{document}